\documentclass[%
reprint,
superscriptaddress,
amsmath,amssymb,
aps,
prb,
floatfix,
]{revtex4-2}

\usepackage{graphicx}
\usepackage{dcolumn}
\usepackage{bm}

\usepackage{color,graphicx}
\usepackage{amsmath}
\usepackage{amssymb}
\usepackage{amsthm}
\usepackage{amsfonts}
\usepackage{braket}
\usepackage[arrowdel]{physics}

\begin{document}
	
	\preprint{APS/123-QED}
	
	\title{Spin dependent analysis of homogeneous and inhomogeneous exciton decoherence in magnetic fields
	}
	\author{V. Laurindo Jr.}
	\affiliation{Departamento de Física, Universidade Federal de São Carlos, 13565-905, São Carlos, SP, Brazil}
	\author{E. D. Guarin Castro}
	\affiliation{Departamento de Física, Universidade Federal de São Carlos, 13565-905, São Carlos, SP, Brazil}
	\author{G. M. Jacobsen}
	\affiliation{Departamento de Física, Universidade Federal de São Carlos, 13565-905, São Carlos, SP, Brazil}
	\author{E. R. C. de Oliveira}
	\affiliation{Departamento de Física, Universidade Federal de São Carlos, 13565-905, São Carlos, SP, Brazil}
	\author{J. F. M. Domenegueti}
	\affiliation{Departamento de Física, Universidade Federal de São Carlos, 13565-905, São Carlos, SP, Brazil}
	\author{B. Alén}
	\affiliation{Instituto de Micro y Nanotecnología, IMN-CNM, CSIC (CEI UAM + CSIC), Tres Cantos, E-28760, Madrid, Spain}
	\author{Yu. I. Mazur}
	\affiliation{Institute for Nanoscience and Engineering, University of Arkansas, Fayetteville, 72701, Arkansas, USA}
	\author{G. J. Salamo}
	\affiliation{Institute for Nanoscience and Engineering, University of Arkansas, Fayetteville, 72701, Arkansas, USA}
	\author{G. E. Marques}
	\affiliation{Departamento de Física, Universidade Federal de São Carlos, 13565-905, São Carlos, SP, Brazil}
	\author{E. Marega Jr.}
	\affiliation{Instituto de Física de São Carlos, Universidade de São Paulo, 13566-590, São Carlos, SP, Brazil}
	\author{ M. D. Teodoro}
	\email{mdaldin@gmail.com}
	\affiliation{Departamento de Física, Universidade Federal de São Carlos, 13565-905, São Carlos, SP, Brazil}
	\author{ V. Lopez-Richard}
	\email{vlopez@df.ufscar.br}
	\affiliation{Departamento de Física, Universidade Federal de São Carlos, 13565-905, São Carlos, SP, Brazil}

	\date{\today}
	
	\begin{abstract}
		This paper discusses the combined effects of optical excitation power, interface roughness, lattice temperature, and applied magnetic fields on the spin-coherence of excitonic states in GaAs/AlGaAs multiple quantum wells. For low optical powers, at lattice temperatures between $4 \text{ K}$ and $50 \text{ K}$, the scattering with acoustic phonons and short-range interactions appear as the main decoherence mechanisms. Statistical fluctuations of the band-gap however become also relevant in this regime and we were able to deconvolute them from the decoherence contributions. The circularly polarized magneto-photoluminescence unveils a non-monotonic tuning of the coherence for one of the spin components at low magnetic fields. This effect has been ascribed to the competition between short-range interactions and spin-flip scattering, modulated by the momentum relaxation time.
	\end{abstract}
	
	
	\maketitle
	
	
	\section{\label{sec:level1}Introduction}
	
	Electronic spin in semiconductors has become a potential building block for applications in spintronics and quantum information technologies.\cite{Wolf2001,Imamoglu1999} This has been studied in a variety of semiconductor platforms ranging from the traditional III-V and II-VI groups to monolayers of transition-metal dichalcogenides~\cite{Ohno1999,Greilich2006,xu2014spin,hao2016} and significant efforts have been devoted to controlling and increasing the spin coherence time.~\cite{Syperek2007,Ullah2016,stockill2016}

	Under a non-resonant optical excitation regime, photogenerated spin carriers subsequently lose energy by scattering processes followed by exciton radiative recombination. This photoluminescence (PL) is mediated by decoherence mechanisms that broaden the exciton linewidth, providing information on the time scale in which the exciton can be coherently manipulated.~\cite{Moody2015} Under an external applied magnetic field, the additional confinement in $xy$\nobreakdash-plane can tune the scattering processes, also increasing the exciton spin relaxation time~\cite{Wang2014} that could lead to spontaneous coherence.~\cite{high2012,Voronova,butov2002,High2008} Therefore, there is an active search for understanding how and under which conditions different decoherence mechanisms of excitons are triggered.  
	To answer these questions we investigated the relaxation process of excitons and the effect of external magnetic fields on the tuning of the spin-coherence in quantum wells (QWs). By reducing the lattice temperature and for low excitation powers, the presence of band-gap fluctuations stabilizes the carriers effective temperature at values higher than the lattice temperature. By applying a magnetic field in this regime, the exciton spin coherence can be tuned. In this case, the combination of short-range interactions and spin-flip scattering is the leading mechanism as supported by our simulations.

	\section{Sample and Experimental Setup}
	A multiple QW heterostructure was grown via molecular-beam epitaxy on an undoped GaAs(100) substrate, consisting of twenty QWs with individual width of $55 \text{ \AA}$ and Al$_{0.36}$Ga$_{0.64}$As barriers with individual thickness of $\sim 300 \text{ \AA}$ - thick enough to avoid carrier tunneling between consecutive wells. Continuous-wave PL measurements were performed at temperatures ranging from $3.6 \text{ K}$ up to $80 \text{ K}$ by means of a confocal microscope with samples placed inside a magneto-optical cryostat (Attocube/Attodry 1000). Magnetic fields up to $6 \text{ T}$ were applied at cryogenic temperatures. A linear polarized diode laser (PicoQuant LDH Series) was used as an excitation source ($\lambda = 730 \text{ nm}$) focused on a spot diameter of $\sim 1 \ \text{\textmu m}$. A set of polarizers was used in order to identify the correspondent sigma plus ($\sigma^{+}$) and minus ($\sigma^{-}$) optical component emissions from the sample, which were magnified into a $\sim 50 \ \text{\textmu m}$ optical fiber acting as a pinhole, dispersed by a $75 \text{ cm}$ spectrometer (Andor / Shamrock), and detected by a Silicon charge-coupled device (Andor / Idus).
	
	\section{Results and Discussion}
\begin{figure}[ht]
		\includegraphics{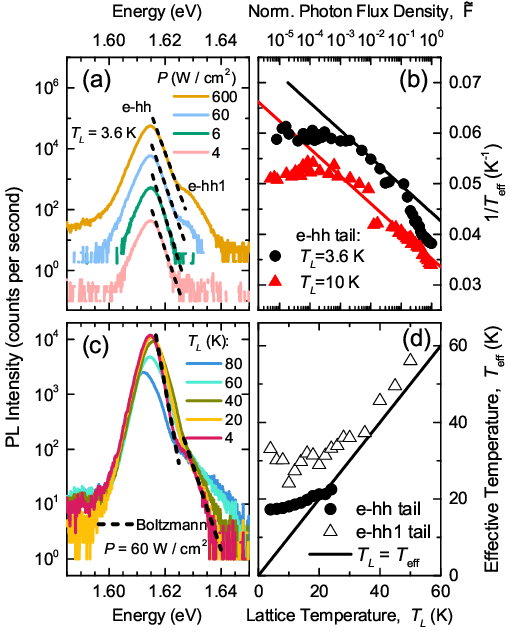}
		\caption{\label{fig1}  (a) PL spectra emission for different laser ($\hbar\omega_{\text{l}}=1.69 \text{ eV}$) power densities at a lattice temperature of $3.6 \text{ K}$. (b) Inverse of the effective temperature obtained from the e-hh spectral tail versus the normalized photon flux density, $\tilde{F}$, at a lattice temperature of $3.6 \text{ K}$ and $10 \text{ K}$. Solid lines simulate the effective temperature variation produced by LO-phonon scattering, $F\propto \exp{\left[ -\hbar\omega_{\text{LO}}/(k_{\text{B}}T_{\text{eff}}) \right]}$. (c) PL spectra for different lattice temperatures at a constant laser power density of $60 \text{ W/cm}^{2}$. The high-energy side of the e-hh and e-hh1 emissions were fitted using Boltzmann functions (dashed lines). (d) Effective temperature obtained from the e-hh and e-hh1 spectral tails as a function of the lattice temperature. Solid line represents the case of carriers thermalization with the lattice ($T_{\text{eff}}=T_{\text{L}}$).}
	\end{figure}

	The PL spectra at a lattice temperature of $T_{\text{L}}=3.6 \text{ K}$ are presented in Fig.~\ref{fig1}(a) for various laser power densities. Here, the main electron-heavy hole emission peak at $1.615 \text{ eV}$, labeled as e-hh, corresponds to the $1s$ ground state recombination.~\cite{Molenkamp1988} At high laser power densities an emission at $1.625\text{ eV}$, labeled as e-hh1, is ascribed to the $2s$ exciton state.~\cite{Molenkamp1988,kajikawa1993}
	
	Within the parabolic band approximation, the intensity of the PL signal, $L$, in quantum wells can be calculated as being proportional to:\cite{bastard1990} 
	\begin{equation}\label{plsign}
		L\propto \bra{u_\text{h}} \mathbf{\epsilon} \cdot \mathbf{p} \ket{u_\text{e}}^2D(\hbar\omega,T_\text{e},T_\text{h}),
	\end{equation}
	where $ \bra{u_\text{h}} \mathbf{\epsilon} \cdot \mathbf{p} \ket{u_\text{e}}^2$ is the dipole matrix element and
	\begin{multline}\label{pls}
		D(\hbar\omega,T_\text{e},T_\text{h}) = \frac{m_\text{r}}{\pi\hbar^2}f^\text{FD}_{\mu_\text{e}}\left(E_\text{g}+E^\text{e}_{1}+\frac{m_\text{r}}{m_\text{e}}\mathcal{E},T_\text{e}\right) \\ f^\text{FD}_{-\mu_\text{h}}\left(E^\text{h}_{1}+\frac{m_\text{r}}{m_\text{h}}\mathcal{E},T_\text{h}\right)H(\mathcal{E}),
	\end{multline}
	being $\hbar\omega$ the emitted photon energy and $f^\text{FD}_{\mu_i}(E,T)$ the Fermi-Dirac distribution function for electron and hole subsystems ($i=$e,h) with effective temperatures $T_\text{e}$ and $T_\text{h}$, and chemical potentials $\mu_\text{e}$ and $-\mu_\text{h}$, respectively. These chemical potentials are measured from the valence band top. The reduced mass is $m_\text{r}=m_\text{e}m_\text{h}/(m_\text{e}+m_\text{h})$, $H (\mathcal{E})$ is the Heaviside step function. and $\mathcal{E} = \hbar\omega - (E_\text{g}+E^\text{e}_{1}+E^\text{h}_{1})$ with $E_\text{g}$ as the band-gap energy and $E^{e(h)}_{1}$ the energy of the ground state for electrons (holes).
	
	In the high-energy side of the emission spectra, the Fermi-Dirac distribution can be approximated to a Boltzmann like function,
	\begin{equation}\label{transBoltz}
		f^\text{FD}_{\mu}(E,T)|_{\frac{E-\mu}{k_\text{B}T}\gg 1} = f^\text{B}(E,T) = \exp \left( -\frac{E- \mu}{k_\text{B}T} \right),
	\end{equation}
	and the quality of this approximation can be assessed using as reference the relative reduction of the distribution function, independently on the position of the chemical potential. It can be readily demonstrated that, the condition 
	\begin{equation}
		\frac{f^\text{B}(E,T)-f^\text{FD}_{\mu}(E,T)}{f^\text{FD}_{\mu}(E,T)}100\% < 10\% ,
	\end{equation}
	is attained once $f^\text{FD}_{\mu}(E,T) < 1/11$. Thus, the relative reduction of at least one order of magnitude below the degenerate condition, where $f^\text{FD}_{\mu}(E,T)=1$, already guarantees staying bellow 10\% discrepancy of the Boltzmann approximation with respect to the Fermi-Dirac distribution. Under this approximation, Eq.\ref{pls} transforms into
	\begin{multline}\label{pls2}
		D(\hbar\omega,T_\text{e},T_\text{h}) = \frac{m_\text{r}}{\pi\hbar^2} \exp\left(-\frac{E_\text{g}+E_{1}^\text{e}+\frac{m_\text{r}}{m_\text{e}}\mathcal{E}-\mu_\text{e}}{k_\text{B}T_\text{e}}\right) \\
		\exp\left(-\frac{E_{1}^\text{h}+\frac{m_\text{r}}{m_\text{h}}\mathcal{E}+\mu_\text{h}}{k_\text{B}T_\text{h}}\right)H(\mathcal{E}),
	\end{multline}
	and so the intensity becomes proportional to 
	\begin{equation}\label{eqboltzmann}
		L(\hbar\omega) \propto \exp\left({-\frac{\hbar\omega}{k_\text{B}T_\text{eff}}}\right),
	\end{equation}
	with $(T_\text{eff})^{-1} \equiv m_\text{r} [(m_\text{e}T_\text{e})^{-1} + (m_\text{h}T_\text{h})^{-1}]$ as the e-h pair effective temperature. A discussion about the nature $T_\text{eff}$ and the sources that contribute to it can be found in Ref.~\cite{edgar2021}. Although this picture points, a priori, to the possibility of the electron and hole subsystems not being mutually thermalized,\cite{bastard1990,edgar2021} there is no reason to assume that in this case, so $T_\text{e}= T_\text{h} = T_\text{eff}$. 
	
	Following the above-presented approximation, the e-hh high-energy spectral tail in Fig.~\ref{fig1}(a) (black dashed lines) can be described by a Boltzmann distribution function, as indicated by Eq.~\ref{eqboltzmann}. The value of $1/T_{\text{eff}}$ extracted from the high-energy tail of the e-hh emission  is presented in Fig.~\ref{fig1}(b) as a function of the photon flux density defined as $F=P/(\hbar\omega_{\text{l}})$, with $P$ being the laser power density and $\hbar\omega_{\text{l}}$ the laser energy.~\cite{Shah1969} In Fig.~\ref{fig1}(b), the flux density has been normalized to the maximum value of the experimental range, $\tilde{F}=F/F_{\text{max}}$ for two different lattice temperatures. When the longitudinal optical (LO-)phonon scattering is the most efficient energy relaxation process, the relation $F\propto \exp{\left[ -\hbar\omega_{\text{LO}}/(k_{\text{B}}T_{\text{eff}}) \right]}$ is expected, with $\hbar\omega_{\text{LO}}=36.5 \text{ meV}$ being the LO-phonon energy in GaAs.~\cite{Shah1969} This is depicted in Fig.~\ref{fig1}(b) by solid lines, in good agreement with the experimental data at high incident photon fluxes ($\tilde{F}>10^{-3}$). At low optical power densities the effective temperature deviates from this trend, stabilizing at a constant value of $17\text{ K}$ for $T_{\text{L}}=3.6 \text{ K}$, and $19\text{ K}$ for $T_{\text{L}}=10 \text{ K}$. 
	This points to a constriction of the LO-phonon scattering, enhancing the relative contribution of the carrier-carrier interaction which stabilizes the effective temperature.~\cite{Hellmann1994} As demonstrated below, this constriction is triggered in the regime where the energy band-gap fluctuations, caused by roughness at the GaAs/AlGaAs interfaces~\cite{Behrend1996} play a dominant role.~\cite{Srinivas1992}
	
	The PL spectra obtained with a laser power density of $60 \text{ W/cm}^{2}$ are shown in Fig.~\ref{fig1}(c) varying $T_{\text{L}}$. In this case the e-hh1 emission is well resolved and $T_{\text{eff}}$ is displayed in Fig.~\ref{fig1}(d) for both e-hh and the e-hh1 tails as a function of $T_{\text{L}}$. The condition of perfect thermalization with the lattice, $T_{\text{eff}}=T_{\text{L}}$, is also represented. The excited states show higher temperatures than the ground state and by increasing $T_{\text{L}}$ a more efficient thermalization with the lattice is observed. We should note that for temperatures above $25 \text{ K}$, the e-hh tail exhibits a shoulder around $1.620 \text{ eV}$ that hampers the extraction of reliable values of $T_{\text{eff}}$. The stabilization of the effective temperature at lower lattice temperatures indicates a constriction of phonon-mediated relaxation that impedes the thermalization with the lattice.

	\subsection{Interface roughness effect} For assessing the relative role of different decoherence mechanisms, the full width at half maximum (FWHM) of the emission lines can be examined. Although the coherence loss by time-irreversible processes can be mapped by analyzing how the FWHM changes with temperature and external fields, this parameter is also affected by statistical fluctuations of the spatial variation of the light sources (excitons in this case). So the role played by the statistics of the interface roughness must be determined.
	
	\begin{figure*}
		\centering
		\includegraphics{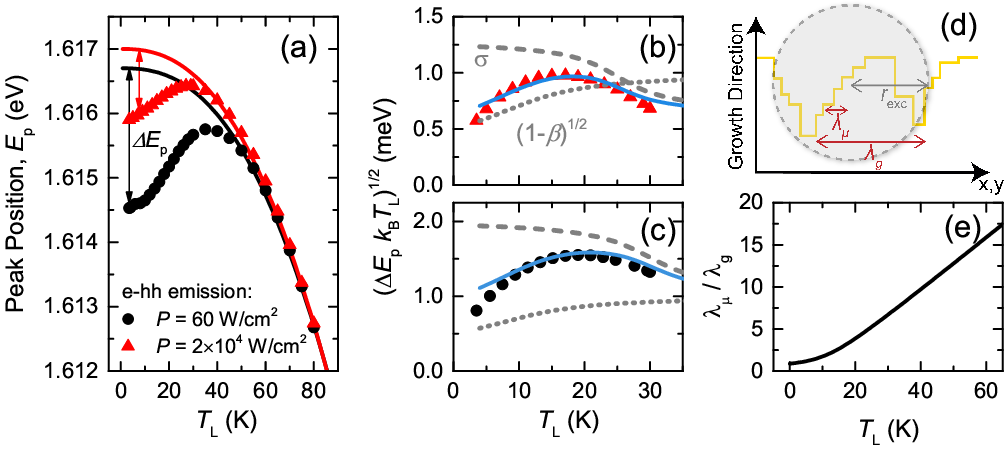}
		\caption{\label{fig2}\textbf(a) PL peak energy, $E_{\text{p}}$, as a function of temperature for laser power densities of $2\times 10^4 \text{W/cm}^{2}$ and $60 \text{ W/cm}^{2}$. Solid curves represent the simulations based on Eq.~\ref{Ep1}. Temperature dependence of $\sqrt{\Delta E_{\text{p}} k_{\text{B}} T_{\text{L}}} \simeq \sigma \sqrt{1-\beta}$ for (b) $P=2\times 10^4 \text{W/cm}^{2}$ and (c) $P=60 \text{ W/cm}^{2}$. Simulations (solid blue lines) were obtained by using step-like functions for $\sigma$ (dashed line) and $\left(1-\beta\right)^{1/2}$ (dotted line). (d) Schematic representation of the characteristic lengths of the band gap fluctuations, $\lambda_{\mu}$ and $\lambda_{\text{g}}$, as compared with the excitonic radius, $r_{\text{exc}}$. (e) Temperature dependence of the $\lambda_{\mu}/\lambda_{\text{g}}$ ratio calculated from the $\beta$ obtained in (b) and (c).}
	\end{figure*}
	
	In order to assess this effect, the lattice temperature dependence of the e-hh peak position, $E_{\text{p}}$, has been displayed in Fig.~\ref{fig2}(a) for two laser power densities along with simulations using the expression~\cite{passler1997}
	\begin{equation}\label{Ep1}
		E_{\text{p}}(T_{\text{L}})=E_{\text{g}}- \frac{\alpha \Theta}{2}\left[\sqrt[p]{1+\left(\frac{2T_{\text{L}}}{\Theta}\right)^p}-1\right]+E^\text{e}_{1}+E^\text{h}_{1},
	\end{equation}
	
	where $E_{\text{g}}=1.519 \text{ eV}$ is the GaAs energy band-gap at $T_{\text{L}}=0 \text{ K}$, and $E^\text{e}_{1}+E^\text{h}_{1}=98\text{ meV}$ is the extra energy given by the QW confinement, $\alpha = 5.405 \times 10^{-4} \text{ eV/K}$,~\cite{Madelung2004} $\Theta \equiv \hbar\omega_{LO}/ k_{B} $ with $\hbar\omega_{LO}=36$ meV for GaAs, while $p=2.5$ for $60 \text{ W/cm}^{2}$ and $p=2.45$ for $2\times 10^4 \text{W/cm}^{2}$. Here, $p$ is a parameter related to the shape of the electron-phonon spectral function.~\cite{passler1997,Teodoro2008}
	The deviation, $\Delta E_{p}$, of the theoretical expectations and the experiment, is produced by local band-gap fluctuations provoked by interface roughness.~\cite{Teodoro2008} At low temperatures, excitons can be trapped into these fluctuations and, by increasing $T_{\text{L}}$, they progressively diffuse and recombine radiatively from higher energy states.~\cite{RUNGE2003149} By increasing the laser power density~\cite{Teodoro2008} the band-gap fluctuations are effectively screened reducing $\Delta E_p$, as confirmed in Figs.~\ref{fig2}(a), (b) and (c). These effects can be analyzed by using the model reported in Ref.~\citenum{Mattheis2007}, which describes the emission intensity as a function of the photon energy $E$, as 
	\begin{multline}\label{Matt1}
		\phi(E)=\text{erfc}\left( \frac{E_g-E+\beta\sigma^2/k_BT_{\text{L}}}{\sqrt{2}\sigma}\right)E^2 \\ \exp \left( -\frac{E-\mu_0-\beta E_g}{k_BT_{\text{L}}} + \frac{\beta^2 \sigma^2}{2 (k_BT_{\text{L}})^2 }\right),
	\end{multline}
	where $\sigma$ is the standard deviation of the local band-gap fluctuations,~\cite{Christen1990,Teodoro2008} and $\beta=[ 1+ (\lambda_\mu/\lambda_g)^2]^{-1/2} \in (0,1]$ depends on the ratio of the characteristic length of the carriers transport, $\lambda_\mu$, with respect to the correlation length scale of the fluctuations, $\lambda_g$, and ponders the trapping efficiency: $\beta \to 0$, small-scale fluctuations (inefficient trapping) and $\beta =1$, large-scale fluctuations (efficient trapping). This has been schematically represented in Fig.~\ref{fig2}(d).
	
	For relatively large arguments of the complementary error function, $\text{erfc}(\xi) \approx \exp(-\xi^2)/\sqrt{\pi}\xi$,~\cite{andrews_1997} and the intensity becomes proportional to  
	\begin{equation}\label{Matt2}
		\phi(E)\propto \exp \left( -\frac{E_g-\frac{\sigma^2}{k_BT_{\text{L}}} (1-\beta)-E}{2\sigma^2} \right).
	\end{equation}
	
	Thus, at low temperatures, its contribution to the FWHM is determined by the standard deviation of the gap fluctuations as $W_\sigma=2\sqrt{2\ln2}\cdot \sigma$, while $\Delta E_p$ can be approximated as
	\begin{equation}\label{Ep2}
		\Delta E_p(T_{\text{L}}) \simeq \frac{\sigma^2}{k_{\text{B}} T_{\text{L}}} (1-\beta).
	\end{equation}
	
\begin{figure}
		\includegraphics{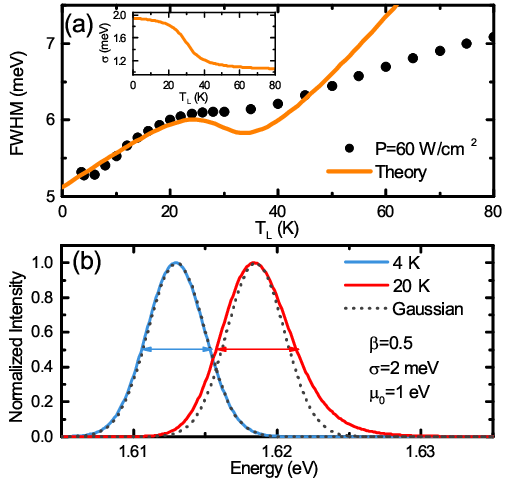}
		\caption{\label{fig3} \textbf (a) FWHM as a function of the lattice temperature for the e-hh emission and ${P} = 60 \text{ W/cm}^{2}$(dots). The simulated FWHM considering homogeneous and inhomogeneous contributions is also displayed (orange line). The $\sigma(T_{L})$ function used for the simulation has been plotted in the inset. (b) Simulated PL spectra with the same $\sigma$ and $\beta$ for $T_{\text{L}}=4~\text{ K}$ (blue line) and $T_{\text{L}}=10~\text{ K}$ (red line), obtained using Eq.~\ref{Matt1} and Gaussian functions (dashed lines) according to Eq.~\ref{Matt2}.}
	\end{figure}

	After extracting the difference between Eq.~\ref{Ep1} and the experimental data, the value of $\sqrt{\Delta E_p(T_{\text{L}}) k_{\text{B}} T_{\text{L}}}=\sigma \sqrt{1-\beta}$ has been displayed in Figs.~\ref{fig2}(b) and c for $P=2\times 10^4 \text{ W/cm}^{2}$ and $P=60 \text{ W/cm}^{2}$, respectively, showing a nonmonotonic behavior between 4 and 50 K. The corresponding experimental values of FWHM have been obtained by a Gaussian fitting of the e-hh emission as a function of $T_{\text{L}}$ for the case of $P=60 \text{ W/cm}^{2}$ and displayed in Fig.~\ref{fig3}(a). In this case a bump appears in the temperature range where the gap fluctuation effects are more evident. Since both power and temperature affect the way the fluctuations are screened, they can also tune the effective values of $\sigma$ and $\beta$. Increasing power and/or temperature provokes an apparent homogenization of the fluctuations (through screening), reducing $\sigma$ while favoring exciton diffusion that translates into a $\beta$ decrease. By assuming $\sigma$ and $\beta$ as soft step-like functions with maximal $\sigma$ and $\beta$ for $T_{\text{L}} \to 0$, then $\sigma$ (dashed lines) decreases by increasing $T_{\text{L}}$ while $\sqrt{1-\beta}$ (dotted lines) grows, as illustrated in Figs.~\ref{fig2}(b) and (c). The product $\sigma \sqrt{1-\beta}$ is also represented (blue solid line) reproducing the nonmonotonic behavior of $\sqrt{\Delta E_p(T_{\text{L}}) k_{\text{B}} T_{\text{L}}}$ up to temperatures between 40 and 50~K. The function used to emulate $\beta$ dependence on temperature has been the same in panels b and c of Fig.~\ref{fig2} and this allows extracting the expected ratio $\lambda_\mu/\lambda_g$ that was depicted in panel e. This suggests an increased detrapping as the temperature grows.
	
	In the case of the FWHM, displayed in Fig.~\ref{fig3}(a), the analysis of the temperature modulation must also include the tuning of the homogeneous lifetime broadening. Thus, assuming the homogeneous broadening as a Lorentzian width, $W_h$, and the inhomogeneous fluctuations characterized by a Gaussian width, $W_\sigma=2\sqrt{2\ln2}\cdot \sigma$, their relative contribution to the FWHM can be approximated as a Voigt convolution,~\cite{Whiting1968, Olivero1977}
	\begin{equation}\label{fwhmger}
		\text{FWHM}=\frac{W_H}{2} + \sqrt{\frac{W_H^2}{4}+W_\sigma^2}.
	\end{equation}
	
	The homogeneous broadening can be simulated by considering different independent and additive mechanisms~\cite{Srinivas1992} that, for low temperatures, can be reduced to two, 
	\begin{equation}\label{homogelow}
		W_H = \Gamma_{0} +  \Gamma_{\text{LA}},
	\end{equation}
	where $\Gamma_{\text{0}}$ represents the intrinsic 2D-excitonic linewidth, associated to exciton-exciton and defects scattering;~\cite{Srinivas1992,Hellmann1994} and $ \Gamma_{\text{LA}} = 2\gamma_{\alpha} T_{\text{L}}$ arises from LA-phonon scattering. Using the function $\sigma$ displayed in the inset of Fig.~\ref{fig3}(a), the best simulation of FWHM at low temperatures was obtained with $\Gamma_{0}=1.0 \text{ meV}$ and $\gamma_{\alpha}=0.045 \text{ meV/K}$, as displayed in Fig.~\ref{fig3}(a). Note that the simulation and the experimental values disagree for higher temperatures. As displayed in Fig.~\ref{fig3}(b), by comparing the results using Eq.~\ref{Matt1} and the approximation of Eq.~\ref{Matt2}, this latter does not account for the whole characterization of the spectral width modulation with temperature. A small but discernible modulation, for constant $\sigma$, already occurs at low $T_\text{L}$. As predicted in Ref.~\citenum{lee1986FWHM}, the LO-phonon scattering and ionized impurities interaction do not significantly contribute to the broadening of the linewidth within the temperature range analyzed. For $T_\text{L}<40\text{ K}$, band-gap fluctuations broadening dominates and responds for the observed bump that can be ascribed to their effective homogenization as $\sigma$ decreases with $T_{\text{L}}$.
	
	\subsection {Magnetic fields effects}
\begin{figure}
		\includegraphics{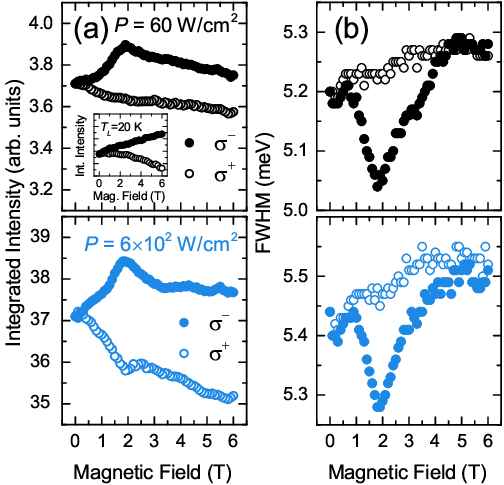}
		\caption{\label{fig4}\textbf Magnetic field dependence of: (a) integrated intensity, and (b) FWHM of the e-hh emission at $T_{\text{L}}=3.6\text{ K}$, obtained with $P=60  \text{ W/cm}^{2}$ (top panels) and $P=6\times 10^2 \text{ W/cm}^{2}$ (bottom panels). Solid and open circles represent the responses measured from $\sigma^-$ and $\sigma^+$ optical component emissions, respectively. The inset in (a) shows the integrated intensity observed at $T_{\text{L}}=20\text{ K}$.}
	\end{figure}

At cryogenic temperatures, phonon-assisted decoherence mechanisms are constricted. This opens the opportunity for studying the scattering processes in the presence of magnetic fields that tune the exciton size~\cite{Aksenov1995,Bansal2007,Chen2020} and the exciton spin relaxation~\cite{Glazov,Wang2014} allowing for additional modulation of the exciton coherence. Figs.~\ref{fig4}(a) and (b) display, respectively, the integrated intensity and FWHM of the e-hh $\sigma^+$ and $\sigma^-$ optical components at $T_\text{L}=3.6\text{ K}$, for $P=60 \text{ W/cm}^{2}$ (top panels) and $P=6\times 10^2 \text{ W/cm}^{2}$ (bottom panels).
	
	The $\sigma^+$ component presents a monotonic dependence with magnetic field for both optical properties in Fig.~\ref{fig4}. By increasing the magnetic field, the $\sigma^+$ exciton population diminishes, as shown in Fig.~\ref{fig4}(a), whereas the FWHM increases in Figs.~\ref{fig4}(b). This FWHM increment is expected from the short-range interaction model, according to which, $\Gamma \propto \sqrt{B}$.~\cite{Ando1974} In contrast, the $\sigma^-$ component in Fig.~\ref{fig4} exhibits a peak response in the integrated intensity and a dip in the FWHM near $B=1.8 \text{ T}$. The peak in the $\sigma^-$ exciton population disappears for $T_\text{L} \geq 20\text{ K}$, as displayed by the inset in Fig. \ref{fig4}(a) for $P=60 \text{ W/cm}^{2}$.
	
\begin{figure}
		\centering
		\includegraphics{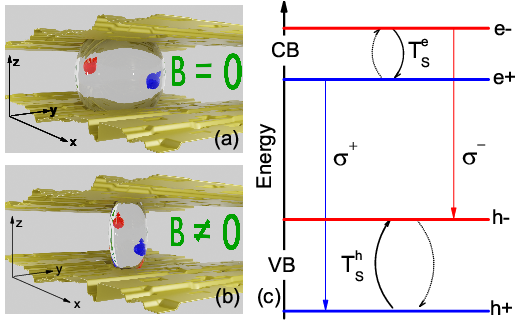}
		\caption{\label{fig5} Illustration of the potential fluctuations caused by roughness along with the GaAs/AlGaAs interfaces and an electron-hole pair confined in the QW with ellipsoids symbolizing the exciton size at (a) $B = 0$ and (b) $B \neq 0$. (c) Representation of the spin relaxation dynamics and the optical recombination selection rules in the presence of a magnetic field.}
	\end{figure}

	The applied magnetic field induces the in-plane confinement of excitons, which leads to a shrinkage of the excitonic wave function and reduces the overlap with larger in-plane disorders as represented schematically in Figs.~\ref{fig5}(a) and (b) for $B =  0 \text{ T}$ and $B \neq  0 \text{ T}$, respectively. The fluctuations length-scale at the interfaces can reach widths of up to $300 \text{ \AA}$ and depths up to $2.8 \text{ \AA}$,\cite{Behrend1996} while the excitonic Bohr radius in bulk GaAs ranges from 160 to 92 \text{\AA} for magnetic fields between zero and $10 \text{ T}$.~\cite{harrison2001,Stepnicki2015} Some studies have described the modulation of the linewidth as an inhomogeneous effect where interface fluctuations of two contrasting scales are averaged over the shrinking exciton radius.~\cite{Bansal2007,harrison2016} Although plausible, they cannot explain the spin modulation resolved in our experiments once the magnetic field modulation of the spin-resolved homogeneous contribution has been overlooked.

	In 1991, by assuming that fluctuations caused by disorder are the dominant contribution to the line broadening, Mena et al. reported a model in Ref.\cite{mena1991} that extended previous theories in the absence of magnetic fields \cite{goede1978,singh1984,schubert1984,singh1986} to the magnetic field modulation. They ultimately concluded that the application of the magnetic field would enhance the value of the linewidth as the exciton radius is reduced and the disorder averaging shrinks (this is represented in panels (a) and (b) of Fig.\ref{fig5}). However, this has been ever since in contradiction with early measurements \cite{sakaki1985,vahala1987} and with the experimental observation of Ref.\cite{Bansal2007} where the linewidth decreases as a function of the magnetic field for 5-10~nm QWs. Although no modeling was provided to reproduce the observations in Refs.\cite{sakaki1985,vahala1987}, they do not preclude the homogeneous broadening tuning with the magnetic field of affecting the FWHM.  In 1995, Aksenov et al. reported in Ref.\cite{Aksenov1995}, an attempt to deconvolute the homogeneous from the inhomogeneous contribution to the nonmonotonic magnetic-field modulation of the FWHM, observed in the PL emission of a GaAs QW. Here, the authors concluded, based on a phenomenological model, that homogeneous effects could be dominant at lower fields. 
	
	We should note that the predictions of the theory presented in Ref.\cite{mena1991} were used to explain the monotonic growth of the FWHM with magnetic field finally observed in Ref.\cite{oliveira1999} (for fields up to 12~T), and also in Ref.\cite{POLIMENI2002} (for fields up to 42~T). Yet, these two papers neglect, a priori, any contribution of homogeneous lifetime modulation at such high fields. Later, in Ref.\cite{Bansal2007}, a plausible mechanism for the inhomogeneous linewidth modulation with magnetic field was introduced, by considering the coexistence of two contrasting length scales. According to that, it is expected that the inhomogeneous linewidth, provoked by larger scale composition fluctuations, decreases with magnetic field (at least for fields up to $\sim$15~T) for 5-10 nm GaAs QWs, following the law  
	\begin{equation}\label{inhomobroader}
		\sigma(B) = \sigma_{0}\frac{1-\exp\left[-R(B)/r_\text{f}\right]}{1-\exp\left[-R_0/r_\text{f}\right]},
	\end{equation}
	where $R(B)$ and $R_0$ are the excitonic radius with and without magnetic field, respectively, $r_\text{f}$ is the effective radius of the large-scale fluctuation in the plane of the QW, and $\sigma_{0}$ is the linewidth at zero magnetic fields. Here, $R(B)$ can be estimated, according to Ref.\cite{piotr2015} as,
	\begin{equation}\label{bohrradius}
		R(B) = \frac{\sqrt{2}R_0}{\left\{1+\left[1+\frac{3}{2}\left(\frac{e^{2}R^4_0 B^{2}}{\hbar^{2}}\right)\right]^{\frac{1}{2}}\right\}^\frac{1}{2}}.
	\end{equation}
	
	Note that, as expected, this magnetic field modulation leads to a monotonic decrease of $\sigma(B)$ as the field grows, shrinking $R(B)$ independently on the spin polarization. In this case, the exciton size shrinkage with magnetic field, represented in Figs. \ref{fig5} (a) and (b), reduces the contact with larger length fluctuating interfaces \cite{Bansal2007}. However, this is still in contradiction with the modulation of the FWHM with magnetic field observed here, that points to the need of introducing additional spin-dependent effects related to the homogeneous contributions to the linewidth.

	In order to account for the polarization resolved modulation of the FWHM, we must consider the electron and hole spin splitting structure and the optical selection rules, as represented in Fig.\ref{fig5} (c). Note, in this picture, that the spin coherence has been assumed to be broken independently in both the conduction and valence bands ground states at rates determined by $1/T^e_{S}$ and  $1/T^{h}_{S}$, respectively. In the case of uncoupled electron-hole pairs, we can approximate their homogeneous lifetime broadening, $W_\text{H}$, as the sum of each component,
	\begin{equation}\label{homogelife}
		W_\text{H}^{\pm} = W_\text{e}^{\pm}+W_\text{h}^{\pm},
	\end{equation}
	where the homogeneous broadening for each component can be expressed as,
	\begin{equation}\label{broden}
		W^{\pm}_i = \sqrt{\frac{2}{\pi}\hbar\omega^i_{c}\frac{\hbar}{\tau^i_{p}}} + \frac{\Gamma^i_{0}}{1 + \left(\frac{\hbar\omega^i_{L}-\hbar\omega^i_{c}}{\hbar/\tau^i_{p}}\right)^{2}}\eta^{\pm}_i,
	\end{equation}
	with $\hbar\omega^i_{c} = \hbar eB/m_ic$, $\hbar\omega^i_{L} = g_i\hbar eB/2m_i c$, and $i=$e (h) labeling electrons (holes) parameters; $m_i$ is the cyclotron effective mass, $g_i$ the Landé factor, and $\tau^i_{p}$ the momentum relaxation time. The first term in Eq.~\ref{broden} corresponds to the contribution of short range scattering~\cite{Ando1974} while the second term considers the spin relaxation \cite{wilam2004,Glazov} inversely proportional to the spin-flip time, $T^i_{S}$, represented in Fig.~\ref{fig5} (c). The spin-flip rates are weighted by the thermal factor $\eta^{\pm}_i = F(E^{\pm}_i- E^{\mp}_i)$ where 
	\begin{equation}
		F(x) = \begin{cases}
			\exp(-\frac{x}{k_\text{B}T}),   & \text{$x \geq 0$}\\
			1, & \text{$x < 0$}
		\end{cases}
	\end{equation}
	which considers the spin energy splitting and the g-factor sign. Within this uncoupled electron-hole pair configuration, the exciton dynamics can be simplified to
	\begin{equation} \label{excitondyna}
		0 = P - \frac{n^{\pm}W^{\pm}_{H}}{2\hbar}-\frac{n^{\pm}}{\tau_{0}},
	\end{equation}
	where $P$ is the electron-hole pair generation rate through illumination, $n^{\pm}$ is the exciton density, and $\tau_{0}$ represents the optical recombination time that leads to 
	\begin{equation}
		n^{\pm} = \frac{2\hbar P}{2\hbar+\tau_{0} W^{\pm}_{h}}.
	\end{equation}
	Thus, the degree of circular polarization (DCP), defined as $DCP = (n^{+}-n^{-})/(n^{+}+n^{-})$, becomes 
	\begin{equation}\label{dcpeq}
		DCP = \frac{W^{-}_{H}-W^{+}_{H}}{\frac{4\hbar}{\tau_{0}}+W^{-}_{H}+W_{H}^{+}}
	\end{equation}
	
	For a quantitative analysis, the actual values of the g-factors, effective masses, and the size of the lateral (in-plane) localization of the lateral movement have been emulated within the parabolic band approximation by using the effective mass Hamiltonian~\cite{llorens2019},
	\begin{equation}\label{hamiltonian}
		H^i_{\pm} = \frac{\hbar^{2}k^{2}}{2m_i}+a_{1}\rho^{2} \pm \frac{1}{2}g^{*}_i\mu_\text{B}B
	\end{equation}
	for both electrons (e) and holes (h). Here, $\rho^{2} = x^{2}+y^{2}$, $\mu_\text{B}$ is the Bohr magneton, and $\mathbf{k} = -i\nabla + \mathbf{A} e/\hbar$, with $\mathbf{A} = B/(2\rho) \hat{\phi})$. The parameter, $a_{1}$, defines the strength of the in-plane confinement of the wavefunction within a site of an effective radius $r^{2}_\text{f} = \hbar/\sqrt{2a_{1}m_{0}}$ and $m_{0}$ is the bare electron mass. The ground state eigenenergy, in this case, is given by~\cite{llorens2019},
	\begin{equation}\label{groundstate}
		E^i_{\pm} = \sqrt{\frac{\hbar^{4}}{m_i r^{4}_\text{f}}+\left(\frac{\hbar \omega^i_{c}}{2}\right)^{2}}\pm \frac{1}{2}g^{*}_i \mu_\text{B} B.
	\end{equation}
	
	\begin{figure}
		\centering
		\includegraphics{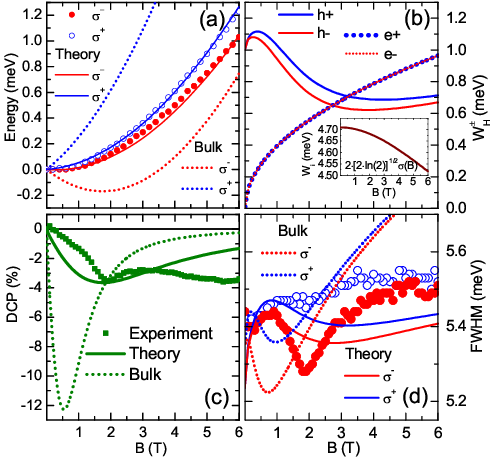}
		\caption{\label{fig6} (a) Experimental peak position shift as a function of the magnetic field strength for $\sigma^+$ (blue circles) and $\sigma^-$ (red dots) polarizations. (b) Calculated homogeneous broadening for both electrons and holes as functions of the magnetic field. The inset shows the inhomogeneous  contribution. (c) Experimental degree of circular polarization (green squares). (d) Experimental values of the FWHM extracted from both circularly polarized emissions as functions of the magnetic field. In all panels, the corresponding emulated values using GaAs QW effective mass parameters are represented with solid curves, whereas those corresponding to bulk GaAs values are represented with dotted lines.}
	\end{figure}

	The experimental electron-hole pair energy peak modulation with magnetic field, relative to the zero field value has been displayed in Fig. \ref{fig6}(a) for $\sigma^+$ (blue circles) and $\sigma^-$ (red dots). The data have been compared to the theoretical electron-hole pair energy calculated from Eq.~\ref{groundstate} as $E_{\pm}^\text{e}+E^\text{h}_{\pm}$. By using the reported electronic structure parameters for GaAs QWs detailed in Table~\ref{Tabela1}, we have been able to confirm the values of the electron and hole g-factors. This allows assessing the size of the effective radius of the lateral confinement site provoked by the interface fluctuations as being  $r_\text{f}=5.5$~nm, which has been used for the calculation of the inhomogeneous broadening modulation with magnetic field according to Eqs.\ref{homogelife} and \ref{broden}, and displayed in the inset of Fig. \ref{fig6}(b). The calculated energy peak obtained by using just bulk GaAs band parameters (included in Table \ref{Tabela1}) have also been added in Fig. \ref{fig6}(a) as reference (dotted lines) pointing to the relevance of considering the modulation of the effective band parameters with confinement. 
	
	\begin{table}
		\centering
		\begin{small}
			\caption{Value of the parameters used in the proposed model. Those ones extracted from the literature have their source cited.} \label{Tabela1}
			\begin{tabular}{c|c|c}
				\hline
				Parameter              & 5.5 nm GaAs QW & Bulk GaAs\\
				\hline
				$m_\text{e}/m_{0}$             & $0.0760$\cite{huant1992} & $0.0664$\cite{winkler2003}\\
				$m_\text{h}/m_{0}$     & $0.45$\cite{Skolnick1976} & $(\gamma_{1}+\gamma_{2})^{-1}=0.112$\cite{winkler2003} \\
				$g_\text{e}$              & $0$\cite{snelling1991} & $-0.44$\cite{winkler2003} \\
				$g_\text{h}$     & $0.7$ & $ 6.\kappa = 7.2$\cite{winkler2003} \\
				$\sigma_{0}$               & $2.0 \text{ meV}$ & $2.0 \text{ meV}$\\
				$R_{0}$                 & $10\text{ nm}$ & $10 \text{ nm}$\\
				$\Gamma^\text{e}_{0}$       & $0$ & $0$\\
				$\Gamma^\text{h}_{0}$       & $1.0 \text{ meV}$ & $1.0 \text{ meV}$\\
				$\hbar/\tau^\text{e}_{p}$       & $0.16 \text{ meV}$ & $0.16 \text{ meV}$\\
				$\hbar/\tau^\text{h}_{p}$       & $0.4 \text{ meV}$ & $0.4 \text{ meV}$\\
				$\tau_{0}$       & $250 \text{ ps}$\cite{Colocci1990} & $250 \text{ ps}$\\
				\hline
			\end{tabular}
		\end{small}
	\end{table}
	
	The corresponding values of the homogeneous width for each spin components in both conduction and valence band have been displayed in Fig. \ref{fig6}(b). One should note that given the negligible value of the electron g-factor and the very small cyclotron mass, the contribution of the electron spin relaxation term has been neglected and that is the reason why we can assume $\Gamma^\text{e}_{0}=0$. The values of the rest of the parameters were set so as to get a good agreement of the calculated DCP, according to Eq. \ref{dcpeq} with the measured values as displayed in Fig. \ref{fig6}(c).
	
	Once all the parameters are determined, the corresponding FWHM for each spin component can be calculated by introducing the homogeneous contribution from Eqs. \ref{homogelife} and \ref{broden} along with the inhomogeneous term, defined by Eqs. \ref{inhomobroader} and \ref{bohrradius}, into Eq. \ref{fwhmger}. A reasonable agreement has been obtained with clear spin asymmetry of the linewidth in Fig. \ref{fig6}(d). Despite the simplicity of the parabolic approximation for the electronic structure model with relevant valence band effects, a nonmonotonic tuning of the FWHM with the magnetic field has also been obtained. The contrast, displayed in Fig. \ref{fig6}(d), of the expected result obtained by using bulk parameters highlights the sensitivity of this response to the band structure modulation with confinement and magnetic field. Note that, besides the eigenenergies, none of the electronic structure parameters was assumed as being spin dependent. That would lead to extra asymmetries of the homogeneous life-time broadening. Thus, the dip in the FWHM observed for the $\sigma^-$ component is a manifestation of increased coherence with respect to $\sigma^+$ determined by the shorter spin-flip rate described by Eq.~\ref{broden}.

	\section{Conclusions}
	A combined experimental and theoretical study was used to investigate exciton decoherence mechanisms in GaAs/AlGaAs multiple QWs. For temperatures below $40 \text{ K}$ and low optical power densities, the hot carrier relaxation is affected by band-gap fluctuations produced by roughness at the GaAs/AlGaAs interfaces. These fluctuations favor carrier-carrier interactions that stabilize the effective temperature. The PL linewidth was used to characterize the exciton coherence. Both LA-phonon interaction and band-gap fluctuations affect this parameter at cryogenic temperatures. Our results demonstrate a strong modulation of these effects with temperature and optical excitation power that allowed the deconvolution of statistical contributions to the linewidth from actual decoherence mechanisms. To further tune the exciton coherence, a magnetic field was applied. At low optical power densities and lattice temperatures, observations with applied magnetic fields unveiled that spin-flip scattering and short-range interactions become the main decoherence factors responsible for the modulation of the excitonic spin-coherence. We were able to quantitatively evaluate the homogeneous and inhomogeneous contributions to the FWHM dependence with magnetic field. Results show that the homogeneous broadening results from short-range interactions and spin relaxation processes, while the inhomogeneous contribution depends on band-gap fluctuations. Although the latter is responsible for almost 90$\%$ of the FWHM we have demonstrated that it is the homogeneous fraction that induces the spin-asymmetry and most of the non-monotonic modulation with magnetic field.

	\begin{acknowledgments}
		This study was financed in part by the Coordenação de Aperfeiçoamento de Pessoal de Nível Superior - Brasil (CAPES) - Finance Code 001. The authors also acknowledge the financial support of the Fundação de Amparo à Pesquisa do Estado de São Paulo (FAPESP) - grants 2013/18719-1, 2014/19142-2, and 2018/01914-0, the Conselho Nacional de Desenvolvimento Científico e Tecnológico (CNPq), the National Science Foundation (NSF) - grant OIA-1457888. This works was also partially funded by Spanish MICINN under grant PID2019-106088RB-C3 and by the MSCA-ITN-2020 Funding Scheme from the European Union’s Horizon 2020 program under Grant agreement ID: 956548.
	\end{acknowledgments}

	\bibliography{bibliography}

\begin{thebibliography}{59}%
\makeatletter
\providecommand \@ifxundefined [1]{%
 \@ifx{#1\undefined}
}%
\providecommand \@ifnum [1]{%
 \ifnum #1\expandafter \@firstoftwo
 \else \expandafter \@secondoftwo
 \fi
}%
\providecommand \@ifx [1]{%
 \ifx #1\expandafter \@firstoftwo
 \else \expandafter \@secondoftwo
 \fi
}%
\providecommand \natexlab [1]{#1}%
\providecommand \enquote  [1]{``#1''}%
\providecommand \bibnamefont  [1]{#1}%
\providecommand \bibfnamefont [1]{#1}%
\providecommand \citenamefont [1]{#1}%
\providecommand \href@noop [0]{\@secondoftwo}%
\providecommand \href [0]{\begingroup \@sanitize@url \@href}%
\providecommand \@href[1]{\@@startlink{#1}\@@href}%
\providecommand \@@href[1]{\endgroup#1\@@endlink}%
\providecommand \@sanitize@url [0]{\catcode `\\12\catcode `\$12\catcode
  `\&12\catcode `\#12\catcode `\^12\catcode `\_12\catcode `\%12\relax}%
\providecommand \@@startlink[1]{}%
\providecommand \@@endlink[0]{}%
\providecommand \url  [0]{\begingroup\@sanitize@url \@url }%
\providecommand \@url [1]{\endgroup\@href {#1}{\urlprefix }}%
\providecommand \urlprefix  [0]{URL }%
\providecommand \Eprint [0]{\href }%
\providecommand \doibase [0]{https://doi.org/}%
\providecommand \selectlanguage [0]{\@gobble}%
\providecommand \bibinfo  [0]{\@secondoftwo}%
\providecommand \bibfield  [0]{\@secondoftwo}%
\providecommand \translation [1]{[#1]}%
\providecommand \BibitemOpen [0]{}%
\providecommand \bibitemStop [0]{}%
\providecommand \bibitemNoStop [0]{.\EOS\space}%
\providecommand \EOS [0]{\spacefactor3000\relax}%
\providecommand \BibitemShut  [1]{\csname bibitem#1\endcsname}%
\let\auto@bib@innerbib\@empty
\bibitem [{\citenamefont {Wolf}\ \emph {et~al.}(2001)\citenamefont {Wolf},
  \citenamefont {Awschalom}, \citenamefont {Buhrman}, \citenamefont {Daughton},
  \citenamefont {von Moln{\'a}r}, \citenamefont {Roukes}, \citenamefont
  {Chtchelkanova},\ and\ \citenamefont {Treger}}]{Wolf2001}%
  \BibitemOpen
  \bibfield  {author} {\bibinfo {author} {\bibfnamefont {S.~A.}\ \bibnamefont
  {Wolf}}, \bibinfo {author} {\bibfnamefont {D.~D.}\ \bibnamefont {Awschalom}},
  \bibinfo {author} {\bibfnamefont {R.~A.}\ \bibnamefont {Buhrman}}, \bibinfo
  {author} {\bibfnamefont {J.~M.}\ \bibnamefont {Daughton}}, \bibinfo {author}
  {\bibfnamefont {S.}~\bibnamefont {von Moln{\'a}r}}, \bibinfo {author}
  {\bibfnamefont {M.~L.}\ \bibnamefont {Roukes}}, \bibinfo {author}
  {\bibfnamefont {A.~Y.}\ \bibnamefont {Chtchelkanova}},\ and\ \bibinfo
  {author} {\bibfnamefont {D.~M.}\ \bibnamefont {Treger}},\ }\bibfield  {title}
  {\bibinfo {title} {{Spintronics: A Spin-Based Electronics Vision for the
  Future}},\ }\href {https://doi.org/10.1126/science.1065389} {\bibfield
  {journal} {\bibinfo  {journal} {Science}\ }\textbf {\bibinfo {volume}
  {294}},\ \bibinfo {pages} {1488} (\bibinfo {year} {2001})}\BibitemShut
  {NoStop}%
\bibitem [{\citenamefont {Imamoglu}\ \emph {et~al.}(1999)\citenamefont
  {Imamoglu}, \citenamefont {Awschalom}, \citenamefont {Burkard}, \citenamefont
  {DiVincenzo}, \citenamefont {Loss}, \citenamefont {Sherwin},\ and\
  \citenamefont {Small}}]{Imamoglu1999}%
  \BibitemOpen
  \bibfield  {author} {\bibinfo {author} {\bibfnamefont {A.}~\bibnamefont
  {Imamoglu}}, \bibinfo {author} {\bibfnamefont {D.~D.}\ \bibnamefont
  {Awschalom}}, \bibinfo {author} {\bibfnamefont {G.}~\bibnamefont {Burkard}},
  \bibinfo {author} {\bibfnamefont {D.~P.}\ \bibnamefont {DiVincenzo}},
  \bibinfo {author} {\bibfnamefont {D.}~\bibnamefont {Loss}}, \bibinfo {author}
  {\bibfnamefont {M.}~\bibnamefont {Sherwin}},\ and\ \bibinfo {author}
  {\bibfnamefont {A.}~\bibnamefont {Small}},\ }\bibfield  {title} {\bibinfo
  {title} {{Quantum Information Processing Using Quantum Dot Spins and Cavity
  QED}},\ }\href {https://doi.org/10.1103/PhysRevLett.83.4204} {\bibfield
  {journal} {\bibinfo  {journal} {Phys. Rev. Lett.}\ }\textbf {\bibinfo
  {volume} {83}},\ \bibinfo {pages} {4204} (\bibinfo {year}
  {1999})}\BibitemShut {NoStop}%
\bibitem [{\citenamefont {Ohno}\ \emph {et~al.}(1999)\citenamefont {Ohno},
  \citenamefont {Terauchi}, \citenamefont {Adachi}, \citenamefont {Matsukura},\
  and\ \citenamefont {Ohno}}]{Ohno1999}%
  \BibitemOpen
  \bibfield  {author} {\bibinfo {author} {\bibfnamefont {Y.}~\bibnamefont
  {Ohno}}, \bibinfo {author} {\bibfnamefont {R.}~\bibnamefont {Terauchi}},
  \bibinfo {author} {\bibfnamefont {T.}~\bibnamefont {Adachi}}, \bibinfo
  {author} {\bibfnamefont {F.}~\bibnamefont {Matsukura}},\ and\ \bibinfo
  {author} {\bibfnamefont {H.}~\bibnamefont {Ohno}},\ }\bibfield  {title}
  {\bibinfo {title} {{Spin Relaxation in GaAs(110) Quantum Wells}},\ }\href
  {https://doi.org/10.1103/PhysRevLett.83.4196} {\bibfield  {journal} {\bibinfo
   {journal} {Phys. Rev. Lett.}\ }\textbf {\bibinfo {volume} {83}},\ \bibinfo
  {pages} {4196} (\bibinfo {year} {1999})}\BibitemShut {NoStop}%
\bibitem [{\citenamefont {Greilich}\ \emph {et~al.}(2006)\citenamefont
  {Greilich}, \citenamefont {Oulton}, \citenamefont {Zhukov}, \citenamefont
  {Yugova}, \citenamefont {Yakovlev}, \citenamefont {Bayer}, \citenamefont
  {Shabaev}, \citenamefont {Efros}, \citenamefont {Merkulov}, \citenamefont
  {Stavarache}, \citenamefont {Reuter},\ and\ \citenamefont
  {Wieck}}]{Greilich2006}%
  \BibitemOpen
  \bibfield  {author} {\bibinfo {author} {\bibfnamefont {A.}~\bibnamefont
  {Greilich}}, \bibinfo {author} {\bibfnamefont {R.}~\bibnamefont {Oulton}},
  \bibinfo {author} {\bibfnamefont {E.~A.}\ \bibnamefont {Zhukov}}, \bibinfo
  {author} {\bibfnamefont {I.~A.}\ \bibnamefont {Yugova}}, \bibinfo {author}
  {\bibfnamefont {D.~R.}\ \bibnamefont {Yakovlev}}, \bibinfo {author}
  {\bibfnamefont {M.}~\bibnamefont {Bayer}}, \bibinfo {author} {\bibfnamefont
  {A.}~\bibnamefont {Shabaev}}, \bibinfo {author} {\bibfnamefont {A.~L.}\
  \bibnamefont {Efros}}, \bibinfo {author} {\bibfnamefont {I.~A.}\ \bibnamefont
  {Merkulov}}, \bibinfo {author} {\bibfnamefont {V.}~\bibnamefont
  {Stavarache}}, \bibinfo {author} {\bibfnamefont {D.}~\bibnamefont {Reuter}},\
  and\ \bibinfo {author} {\bibfnamefont {A.}~\bibnamefont {Wieck}},\ }\bibfield
   {title} {\bibinfo {title} {{Optical Control of Spin Coherence in Singly
  Charged $(\mathrm{In},\mathrm{Ga})\mathrm{As}/\mathrm{GaAs}$ Quantum Dots}},\
  }\href {https://doi.org/10.1103/PhysRevLett.96.227401} {\bibfield  {journal}
  {\bibinfo  {journal} {Phys. Rev. Lett.}\ }\textbf {\bibinfo {volume} {96}},\
  \bibinfo {pages} {227401} (\bibinfo {year} {2006})}\BibitemShut {NoStop}%
\bibitem [{\citenamefont {Xu}\ \emph {et~al.}(2014)\citenamefont {Xu},
  \citenamefont {Yao}, \citenamefont {Xiao},\ and\ \citenamefont
  {Heinz}}]{xu2014spin}%
  \BibitemOpen
  \bibfield  {author} {\bibinfo {author} {\bibfnamefont {X.}~\bibnamefont
  {Xu}}, \bibinfo {author} {\bibfnamefont {W.}~\bibnamefont {Yao}}, \bibinfo
  {author} {\bibfnamefont {D.}~\bibnamefont {Xiao}},\ and\ \bibinfo {author}
  {\bibfnamefont {T.~F.}\ \bibnamefont {Heinz}},\ }\bibfield  {title} {\bibinfo
  {title} {{Spin and pseudospins in layered transition metal
  dichalcogenides}},\ }\href {https://doi.org/10.1038/nphys2942} {\bibfield
  {journal} {\bibinfo  {journal} {Nat. Phys.}\ }\textbf {\bibinfo {volume}
  {10}},\ \bibinfo {pages} {343} (\bibinfo {year} {2014})}\BibitemShut
  {NoStop}%
\bibitem [{\citenamefont {Hao}\ \emph {et~al.}(2016)\citenamefont {Hao},
  \citenamefont {Moody}, \citenamefont {Wu}, \citenamefont {Dass},
  \citenamefont {Xu}, \citenamefont {Chen}, \citenamefont {Sun}, \citenamefont
  {Li}, \citenamefont {Li}, \citenamefont {MacDonald} \emph
  {et~al.}}]{hao2016}%
  \BibitemOpen
  \bibfield  {author} {\bibinfo {author} {\bibfnamefont {K.}~\bibnamefont
  {Hao}}, \bibinfo {author} {\bibfnamefont {G.}~\bibnamefont {Moody}}, \bibinfo
  {author} {\bibfnamefont {F.}~\bibnamefont {Wu}}, \bibinfo {author}
  {\bibfnamefont {C.~K.}\ \bibnamefont {Dass}}, \bibinfo {author}
  {\bibfnamefont {L.}~\bibnamefont {Xu}}, \bibinfo {author} {\bibfnamefont
  {C.-H.}\ \bibnamefont {Chen}}, \bibinfo {author} {\bibfnamefont
  {L.}~\bibnamefont {Sun}}, \bibinfo {author} {\bibfnamefont {M.-Y.}\
  \bibnamefont {Li}}, \bibinfo {author} {\bibfnamefont {L.-J.}\ \bibnamefont
  {Li}}, \bibinfo {author} {\bibfnamefont {A.~H.}\ \bibnamefont {MacDonald}},
  \emph {et~al.},\ }\bibfield  {title} {\bibinfo {title} {{Direct measurement
  of exciton valley coherence in monolayer ${\mathrm{WSe}}_{2}$}},\ }\href
  {https://doi.org/10.1038/nphys3674} {\bibfield  {journal} {\bibinfo
  {journal} {Nat. Phys.}\ }\textbf {\bibinfo {volume} {12}},\ \bibinfo {pages}
  {677} (\bibinfo {year} {2016})}\BibitemShut {NoStop}%
\bibitem [{\citenamefont {Syperek}\ \emph {et~al.}(2007)\citenamefont
  {Syperek}, \citenamefont {Yakovlev}, \citenamefont {Greilich}, \citenamefont
  {Misiewicz}, \citenamefont {Bayer}, \citenamefont {Reuter},\ and\
  \citenamefont {Wieck}}]{Syperek2007}%
  \BibitemOpen
  \bibfield  {author} {\bibinfo {author} {\bibfnamefont {M.}~\bibnamefont
  {Syperek}}, \bibinfo {author} {\bibfnamefont {D.~R.}\ \bibnamefont
  {Yakovlev}}, \bibinfo {author} {\bibfnamefont {A.}~\bibnamefont {Greilich}},
  \bibinfo {author} {\bibfnamefont {J.}~\bibnamefont {Misiewicz}}, \bibinfo
  {author} {\bibfnamefont {M.}~\bibnamefont {Bayer}}, \bibinfo {author}
  {\bibfnamefont {D.}~\bibnamefont {Reuter}},\ and\ \bibinfo {author}
  {\bibfnamefont {A.~D.}\ \bibnamefont {Wieck}},\ }\bibfield  {title} {\bibinfo
  {title} {{Spin Coherence of Holes in
  $\mathrm{GaAs}/(\mathrm{Al},\mathrm{Ga})\mathrm{As}$ Quantum Wells}},\ }\href
  {https://doi.org/10.1103/PhysRevLett.99.187401} {\bibfield  {journal}
  {\bibinfo  {journal} {Phys. Rev. Lett.}\ }\textbf {\bibinfo {volume} {99}},\
  \bibinfo {pages} {187401} (\bibinfo {year} {2007})}\BibitemShut {NoStop}%
\bibitem [{\citenamefont {Ullah}\ \emph {et~al.}(2016)\citenamefont {Ullah},
  \citenamefont {Gusev}, \citenamefont {Bakarov},\ and\ \citenamefont
  {Hernandez}}]{Ullah2016}%
  \BibitemOpen
  \bibfield  {author} {\bibinfo {author} {\bibfnamefont {S.}~\bibnamefont
  {Ullah}}, \bibinfo {author} {\bibfnamefont {G.~M.}\ \bibnamefont {Gusev}},
  \bibinfo {author} {\bibfnamefont {A.~K.}\ \bibnamefont {Bakarov}},\ and\
  \bibinfo {author} {\bibfnamefont {F.~G.~G.}\ \bibnamefont {Hernandez}},\
  }\bibfield  {title} {\bibinfo {title} {{Long-lived nanosecond spin coherence
  in high-mobility 2DEGs confined in double and triple quantum wells}},\ }\href
  {https://doi.org/10.1063/1.4953007} {\bibfield  {journal} {\bibinfo
  {journal} {J. Appl. Phys.}\ }\textbf {\bibinfo {volume} {119}},\ \bibinfo
  {pages} {215701} (\bibinfo {year} {2016})}\BibitemShut {NoStop}%
\bibitem [{\citenamefont {Stockill}\ \emph {et~al.}(2016)\citenamefont
  {Stockill}, \citenamefont {Le~Gall}, \citenamefont {Matthiesen},
  \citenamefont {Huthmacher}, \citenamefont {Clarke}, \citenamefont {Hugues},\
  and\ \citenamefont {Atatüre}}]{stockill2016}%
  \BibitemOpen
  \bibfield  {author} {\bibinfo {author} {\bibfnamefont {R.}~\bibnamefont
  {Stockill}}, \bibinfo {author} {\bibfnamefont {C.}~\bibnamefont {Le~Gall}},
  \bibinfo {author} {\bibfnamefont {C.}~\bibnamefont {Matthiesen}}, \bibinfo
  {author} {\bibfnamefont {L.}~\bibnamefont {Huthmacher}}, \bibinfo {author}
  {\bibfnamefont {E.}~\bibnamefont {Clarke}}, \bibinfo {author} {\bibfnamefont
  {M.}~\bibnamefont {Hugues}},\ and\ \bibinfo {author} {\bibfnamefont
  {M.}~\bibnamefont {Atatüre}},\ }\bibfield  {title} {\bibinfo {title}
  {{Quantum dot spin coherence governed by a strained nuclear environment}},\
  }\href {https://doi.org/10.1038/ncomms12745} {\bibfield  {journal} {\bibinfo
  {journal} {Nat. Commun.}\ }\textbf {\bibinfo {volume} {7}},\ \bibinfo {pages}
  {1} (\bibinfo {year} {2016})}\BibitemShut {NoStop}%
\bibitem [{\citenamefont {Moody}\ \emph {et~al.}(2015)\citenamefont {Moody},
  \citenamefont {Dass}, \citenamefont {Hao}, \citenamefont {Chen},
  \citenamefont {Li}, \citenamefont {Singh}, \citenamefont {Tran},
  \citenamefont {Clark}, \citenamefont {Xu}, \citenamefont {Bergh{\"a}user}
  \emph {et~al.}}]{Moody2015}%
  \BibitemOpen
  \bibfield  {author} {\bibinfo {author} {\bibfnamefont {G.}~\bibnamefont
  {Moody}}, \bibinfo {author} {\bibfnamefont {C.~K.}\ \bibnamefont {Dass}},
  \bibinfo {author} {\bibfnamefont {K.}~\bibnamefont {Hao}}, \bibinfo {author}
  {\bibfnamefont {C.-H.}\ \bibnamefont {Chen}}, \bibinfo {author}
  {\bibfnamefont {L.-J.}\ \bibnamefont {Li}}, \bibinfo {author} {\bibfnamefont
  {A.}~\bibnamefont {Singh}}, \bibinfo {author} {\bibfnamefont
  {K.}~\bibnamefont {Tran}}, \bibinfo {author} {\bibfnamefont {G.}~\bibnamefont
  {Clark}}, \bibinfo {author} {\bibfnamefont {X.}~\bibnamefont {Xu}}, \bibinfo
  {author} {\bibfnamefont {G.}~\bibnamefont {Bergh{\"a}user}}, \emph {et~al.},\
  }\bibfield  {title} {\bibinfo {title} {{Intrinsic homogeneous linewidth and
  broadening mechanisms of excitons in monolayer transition metal
  dichalcogenides}},\ }\href {https://doi.org/10.1038/ncomms9315} {\bibfield
  {journal} {\bibinfo  {journal} {Nat. Commun.}\ }\textbf {\bibinfo {volume}
  {6}},\ \bibinfo {pages} {1} (\bibinfo {year} {2015})}\BibitemShut {NoStop}%
\bibitem [{\citenamefont {Wang}\ \emph {et~al.}(2014)\citenamefont {Wang},
  \citenamefont {Balocchi}, \citenamefont {Poshakinskiy}, \citenamefont {Zhu},
  \citenamefont {Tarasenko}, \citenamefont {Amand}, \citenamefont {Liu},\ and\
  \citenamefont {Marie}}]{Wang2014}%
  \BibitemOpen
  \bibfield  {author} {\bibinfo {author} {\bibfnamefont {G.}~\bibnamefont
  {Wang}}, \bibinfo {author} {\bibfnamefont {A.}~\bibnamefont {Balocchi}},
  \bibinfo {author} {\bibfnamefont {A.~V.}\ \bibnamefont {Poshakinskiy}},
  \bibinfo {author} {\bibfnamefont {C.~R.}\ \bibnamefont {Zhu}}, \bibinfo
  {author} {\bibfnamefont {S.~A.}\ \bibnamefont {Tarasenko}}, \bibinfo {author}
  {\bibfnamefont {T.}~\bibnamefont {Amand}}, \bibinfo {author} {\bibfnamefont
  {B.~L.}\ \bibnamefont {Liu}},\ and\ \bibinfo {author} {\bibfnamefont
  {X.}~\bibnamefont {Marie}},\ }\bibfield  {title} {\bibinfo {title} {{Magnetic
  field effect on electron spin dynamics in (110) {GaAs} quantum wells}},\
  }\href {https://doi.org/10.1088/1367-2630/16/4/045008} {\bibfield  {journal}
  {\bibinfo  {journal} {New J. Phys.}\ }\textbf {\bibinfo {volume} {16}},\
  \bibinfo {pages} {045008} (\bibinfo {year} {2014})}\BibitemShut {NoStop}%
\bibitem [{\citenamefont {High}\ \emph {et~al.}(2012)\citenamefont {High},
  \citenamefont {Leonard}, \citenamefont {Hammack}, \citenamefont {Fogler},
  \citenamefont {Butov}, \citenamefont {Kavokin}, \citenamefont {Campman},\
  and\ \citenamefont {Gossard}}]{high2012}%
  \BibitemOpen
  \bibfield  {author} {\bibinfo {author} {\bibfnamefont {A.~A.}\ \bibnamefont
  {High}}, \bibinfo {author} {\bibfnamefont {J.~R.}\ \bibnamefont {Leonard}},
  \bibinfo {author} {\bibfnamefont {A.~T.}\ \bibnamefont {Hammack}}, \bibinfo
  {author} {\bibfnamefont {M.~M.}\ \bibnamefont {Fogler}}, \bibinfo {author}
  {\bibfnamefont {L.~V.}\ \bibnamefont {Butov}}, \bibinfo {author}
  {\bibfnamefont {A.~V.}\ \bibnamefont {Kavokin}}, \bibinfo {author}
  {\bibfnamefont {K.~L.}\ \bibnamefont {Campman}},\ and\ \bibinfo {author}
  {\bibfnamefont {A.~C.}\ \bibnamefont {Gossard}},\ }\bibfield  {title}
  {\bibinfo {title} {{Spontaneous coherence in a cold exciton gas}},\ }\href
  {https://doi.org/10.1038/nature10903} {\bibfield  {journal} {\bibinfo
  {journal} {Nature}\ }\textbf {\bibinfo {volume} {483}},\ \bibinfo {pages}
  {584} (\bibinfo {year} {2012})}\BibitemShut {NoStop}%
\bibitem [{\citenamefont {Voronova}\ \emph {et~al.}(2018)\citenamefont
  {Voronova}, \citenamefont {Kurbakov},\ and\ \citenamefont
  {Lozovik}}]{Voronova}%
  \BibitemOpen
  \bibfield  {author} {\bibinfo {author} {\bibfnamefont {N.~S.}\ \bibnamefont
  {Voronova}}, \bibinfo {author} {\bibfnamefont {I.~L.}\ \bibnamefont
  {Kurbakov}},\ and\ \bibinfo {author} {\bibfnamefont {Y.~E.}\ \bibnamefont
  {Lozovik}},\ }\bibfield  {title} {\bibinfo {title} {{Bose Condensation of
  Long-Living Direct Excitons in an Off-Resonant Cavity}},\ }\href
  {https://doi.org/10.1103/PhysRevLett.121.235702} {\bibfield  {journal}
  {\bibinfo  {journal} {Phys. Rev. Lett.}\ }\textbf {\bibinfo {volume} {121}},\
  \bibinfo {pages} {235702} (\bibinfo {year} {2018})}\BibitemShut {NoStop}%
\bibitem [{\citenamefont {Butov}\ \emph {et~al.}(2002)\citenamefont {Butov},
  \citenamefont {Lai}, \citenamefont {Ivanov}, \citenamefont {Gossard},\ and\
  \citenamefont {Chemla}}]{butov2002}%
  \BibitemOpen
  \bibfield  {author} {\bibinfo {author} {\bibfnamefont {L.}~\bibnamefont
  {Butov}}, \bibinfo {author} {\bibfnamefont {C.}~\bibnamefont {Lai}}, \bibinfo
  {author} {\bibfnamefont {A.}~\bibnamefont {Ivanov}}, \bibinfo {author}
  {\bibfnamefont {A.}~\bibnamefont {Gossard}},\ and\ \bibinfo {author}
  {\bibfnamefont {D.}~\bibnamefont {Chemla}},\ }\bibfield  {title} {\bibinfo
  {title} {{Towards Bose--Einstein condensation of excitons in potential
  traps}},\ }\href {https://doi.org/10.1038/417047a} {\bibfield  {journal}
  {\bibinfo  {journal} {Nature}\ }\textbf {\bibinfo {volume} {417}},\ \bibinfo
  {pages} {47} (\bibinfo {year} {2002})}\BibitemShut {NoStop}%
\bibitem [{\citenamefont {High}\ \emph {et~al.}(2008)\citenamefont {High},
  \citenamefont {Novitskaya}, \citenamefont {Butov}, \citenamefont {Hanson},\
  and\ \citenamefont {Gossard}}]{High2008}%
  \BibitemOpen
  \bibfield  {author} {\bibinfo {author} {\bibfnamefont {A.~A.}\ \bibnamefont
  {High}}, \bibinfo {author} {\bibfnamefont {E.~E.}\ \bibnamefont
  {Novitskaya}}, \bibinfo {author} {\bibfnamefont {L.~V.}\ \bibnamefont
  {Butov}}, \bibinfo {author} {\bibfnamefont {M.}~\bibnamefont {Hanson}},\ and\
  \bibinfo {author} {\bibfnamefont {A.~C.}\ \bibnamefont {Gossard}},\
  }\bibfield  {title} {\bibinfo {title} {{Control of Exciton Fluxes in an
  Excitonic Integrated Circuit}},\ }\href
  {https://doi.org/10.1126/science.1157845} {\bibfield  {journal} {\bibinfo
  {journal} {Science}\ }\textbf {\bibinfo {volume} {321}},\ \bibinfo {pages}
  {229} (\bibinfo {year} {2008})}\BibitemShut {NoStop}%
\bibitem [{\citenamefont {Molenkamp}\ \emph {et~al.}(1988)\citenamefont
  {Molenkamp}, \citenamefont {Bauer}, \citenamefont {Eppenga},\ and\
  \citenamefont {Foxon}}]{Molenkamp1988}%
  \BibitemOpen
  \bibfield  {author} {\bibinfo {author} {\bibfnamefont {L.~W.}\ \bibnamefont
  {Molenkamp}}, \bibinfo {author} {\bibfnamefont {G.~E.~W.}\ \bibnamefont
  {Bauer}}, \bibinfo {author} {\bibfnamefont {R.}~\bibnamefont {Eppenga}},\
  and\ \bibinfo {author} {\bibfnamefont {C.~T.}\ \bibnamefont {Foxon}},\
  }\bibfield  {title} {\bibinfo {title} {{Exciton binding energy in (Al,Ga)As
  quantum wells: Effects of crystal orientation and envelope-function
  symmetry}},\ }\href {https://doi.org/10.1103/PhysRevB.38.6147} {\bibfield
  {journal} {\bibinfo  {journal} {Phys. Rev. B}\ }\textbf {\bibinfo {volume}
  {38}},\ \bibinfo {pages} {6147} (\bibinfo {year} {1988})}\BibitemShut
  {NoStop}%
\bibitem [{\citenamefont {Kajikawa}(1993)}]{kajikawa1993}%
  \BibitemOpen
  \bibfield  {author} {\bibinfo {author} {\bibfnamefont {Y.}~\bibnamefont
  {Kajikawa}},\ }\bibfield  {title} {\bibinfo {title} {{Comparison of 1s-2s
  exciton-energy splittings between (001) and (111)
  GaAs/${\mathrm{Al}}_{\mathit{x}}$
  ${\mathrm{Ga}}_{1\mathrm{\ensuremath{-}}\mathit{x}}$As quantum wells}},\
  }\href {https://doi.org/10.1103/PhysRevB.48.7935} {\bibfield  {journal}
  {\bibinfo  {journal} {Phys. Rev. B}\ }\textbf {\bibinfo {volume} {48}},\
  \bibinfo {pages} {7935} (\bibinfo {year} {1993})}\BibitemShut {NoStop}%
\bibitem [{\citenamefont {Bastard}(1990)}]{bastard1990}%
  \BibitemOpen
  \bibfield  {author} {\bibinfo {author} {\bibfnamefont {G.}~\bibnamefont
  {Bastard}},\ }\href@noop {} {\emph {\bibinfo {title} {Wave Mechanics Applied
  to Semiconductor Heterostructures}}}\ (\bibinfo  {publisher} {les éditions
  de physique},\ \bibinfo {address} {Paris},\ \bibinfo {year}
  {1990})\BibitemShut {NoStop}%
\bibitem [{\citenamefont {Guarin~Castro}\ \emph {et~al.}(2021)\citenamefont
  {Guarin~Castro}, \citenamefont {Pfenning}, \citenamefont {Hartmann},
  \citenamefont {Knebl}, \citenamefont {Daldin~Teodoro}, \citenamefont
  {Marques}, \citenamefont {Höfling}, \citenamefont {Bastard},\ and\
  \citenamefont {Lopez-Richard}}]{edgar2021}%
  \BibitemOpen
  \bibfield  {author} {\bibinfo {author} {\bibfnamefont {E.~D.}\ \bibnamefont
  {Guarin~Castro}}, \bibinfo {author} {\bibfnamefont {A.}~\bibnamefont
  {Pfenning}}, \bibinfo {author} {\bibfnamefont {F.}~\bibnamefont {Hartmann}},
  \bibinfo {author} {\bibfnamefont {G.}~\bibnamefont {Knebl}}, \bibinfo
  {author} {\bibfnamefont {M.}~\bibnamefont {Daldin~Teodoro}}, \bibinfo
  {author} {\bibfnamefont {G.~E.}\ \bibnamefont {Marques}}, \bibinfo {author}
  {\bibfnamefont {S.}~\bibnamefont {Höfling}}, \bibinfo {author}
  {\bibfnamefont {G.}~\bibnamefont {Bastard}},\ and\ \bibinfo {author}
  {\bibfnamefont {V.}~\bibnamefont {Lopez-Richard}},\ }\bibfield  {title}
  {\bibinfo {title} {Optical mapping of nonequilibrium charge carriers},\
  }\href {https://doi.org/10.1021/acs.jpcc.1c02173} {\bibfield  {journal}
  {\bibinfo  {journal} {The Journal of Physical Chemistry C}\ }\textbf
  {\bibinfo {volume} {125}},\ \bibinfo {pages} {14741} (\bibinfo {year}
  {2021})},\ \Eprint
  {https://arxiv.org/abs/https://doi.org/10.1021/acs.jpcc.1c02173}
  {https://doi.org/10.1021/acs.jpcc.1c02173} \BibitemShut {NoStop}%
\bibitem [{\citenamefont {Shah}\ and\ \citenamefont {Leite}(1969)}]{Shah1969}%
  \BibitemOpen
  \bibfield  {author} {\bibinfo {author} {\bibfnamefont {J.}~\bibnamefont
  {Shah}}\ and\ \bibinfo {author} {\bibfnamefont {R.~C.~C.}\ \bibnamefont
  {Leite}},\ }\bibfield  {title} {\bibinfo {title} {{Radiative Recombination
  from Photoexcited Hot Carriers in GaAs}},\ }\href
  {https://doi.org/10.1103/PhysRevLett.22.1304} {\bibfield  {journal} {\bibinfo
   {journal} {Phys. Rev. Lett.}\ }\textbf {\bibinfo {volume} {22}},\ \bibinfo
  {pages} {1304} (\bibinfo {year} {1969})}\BibitemShut {NoStop}%
\bibitem [{\citenamefont {Hellmann}\ \emph {et~al.}(1994)\citenamefont
  {Hellmann}, \citenamefont {Koch}, \citenamefont {Feldmann}, \citenamefont
  {Cundiff}, \citenamefont {Göbel}, \citenamefont {Yakovlev}, \citenamefont
  {Waag},\ and\ \citenamefont {Landwehr}}]{Hellmann1994}%
  \BibitemOpen
  \bibfield  {author} {\bibinfo {author} {\bibfnamefont {R.}~\bibnamefont
  {Hellmann}}, \bibinfo {author} {\bibfnamefont {M.}~\bibnamefont {Koch}},
  \bibinfo {author} {\bibfnamefont {J.}~\bibnamefont {Feldmann}}, \bibinfo
  {author} {\bibfnamefont {S.}~\bibnamefont {Cundiff}}, \bibinfo {author}
  {\bibfnamefont {E.}~\bibnamefont {Göbel}}, \bibinfo {author} {\bibfnamefont
  {D.}~\bibnamefont {Yakovlev}}, \bibinfo {author} {\bibfnamefont
  {A.}~\bibnamefont {Waag}},\ and\ \bibinfo {author} {\bibfnamefont
  {G.}~\bibnamefont {Landwehr}},\ }\bibfield  {title} {\bibinfo {title}
  {{Dephasing of excitons in a
  ${\mathrm{CdTe}}/{\mathrm{Cd}}_{0.86}{\mathrm{Mn}}_{0.14}{\mathrm{Te}}$
  multiple quantum well}},\ }\href
  {https://doi.org/https://doi.org/10.1016/0022-0248(94)90908-3} {\bibfield
  {journal} {\bibinfo  {journal} {J. Cryst. Growth}\ }\textbf {\bibinfo
  {volume} {138}},\ \bibinfo {pages} {791 } (\bibinfo {year}
  {1994})}\BibitemShut {NoStop}%
\bibitem [{\citenamefont {Behrend}\ \emph {et~al.}(1996)\citenamefont
  {Behrend}, \citenamefont {Wassermeier}, \citenamefont {Braun}, \citenamefont
  {Krispin},\ and\ \citenamefont {Ploog}}]{Behrend1996}%
  \BibitemOpen
  \bibfield  {author} {\bibinfo {author} {\bibfnamefont {J.}~\bibnamefont
  {Behrend}}, \bibinfo {author} {\bibfnamefont {M.}~\bibnamefont
  {Wassermeier}}, \bibinfo {author} {\bibfnamefont {W.}~\bibnamefont {Braun}},
  \bibinfo {author} {\bibfnamefont {P.}~\bibnamefont {Krispin}},\ and\ \bibinfo
  {author} {\bibfnamefont {K.~H.}\ \bibnamefont {Ploog}},\ }\bibfield  {title}
  {\bibinfo {title} {{Formation of GaAs/AlAs(001) interfaces studied by
  scanning tunneling microscopy}},\ }\href
  {https://doi.org/10.1103/PhysRevB.53.9907} {\bibfield  {journal} {\bibinfo
  {journal} {Phys. Rev. B}\ }\textbf {\bibinfo {volume} {53}},\ \bibinfo
  {pages} {9907} (\bibinfo {year} {1996})}\BibitemShut {NoStop}%
\bibitem [{\citenamefont {Srinivas}\ \emph {et~al.}(1992)\citenamefont
  {Srinivas}, \citenamefont {Hryniewicz}, \citenamefont {Chen},\ and\
  \citenamefont {Wood}}]{Srinivas1992}%
  \BibitemOpen
  \bibfield  {author} {\bibinfo {author} {\bibfnamefont {V.}~\bibnamefont
  {Srinivas}}, \bibinfo {author} {\bibfnamefont {J.}~\bibnamefont
  {Hryniewicz}}, \bibinfo {author} {\bibfnamefont {Y.~J.}\ \bibnamefont
  {Chen}},\ and\ \bibinfo {author} {\bibfnamefont {C.~E.~C.}\ \bibnamefont
  {Wood}},\ }\bibfield  {title} {\bibinfo {title} {{Intrinsic linewidths and
  radiative lifetimes of free excitons in GaAs quantum wells}},\ }\href
  {https://doi.org/10.1103/PhysRevB.46.10193} {\bibfield  {journal} {\bibinfo
  {journal} {Phys. Rev. B}\ }\textbf {\bibinfo {volume} {46}},\ \bibinfo
  {pages} {10193} (\bibinfo {year} {1992})}\BibitemShut {NoStop}%
\bibitem [{\citenamefont {Pässler}(1997)}]{passler1997}%
  \BibitemOpen
  \bibfield  {author} {\bibinfo {author} {\bibfnamefont {R.}~\bibnamefont
  {Pässler}},\ }\bibfield  {title} {\bibinfo {title} {{Basic Model Relations
  for Temperature Dependencies of Fundamental Energy Gaps in Semiconductors}},\
  }\href
  {https://doi.org/https://doi.org/10.1002/1521-3951(199703)200:1<155::AID-PSSB155>3.0.CO;2-3}
  {\bibfield  {journal} {\bibinfo  {journal} {Phys. Status Solidi B}\ }\textbf
  {\bibinfo {volume} {200}},\ \bibinfo {pages} {155} (\bibinfo {year}
  {1997})}\BibitemShut {NoStop}%
\bibitem [{\citenamefont {Madelung}(2004)}]{Madelung2004}%
  \BibitemOpen
  \bibfield  {author} {\bibinfo {author} {\bibfnamefont {O.}~\bibnamefont
  {Madelung}},\ }\href@noop {} {\emph {\bibinfo {title} {Semiconductors: Data
  handbook (CD-ROM)}}},\ \bibinfo {edition} {3rd}\ ed.\ (\bibinfo  {publisher}
  {Springer},\ \bibinfo {year} {2004})\BibitemShut {NoStop}%
\bibitem [{\citenamefont {Teodoro}\ \emph {et~al.}(2008)\citenamefont
  {Teodoro}, \citenamefont {Dias}, \citenamefont {Laureto}, \citenamefont
  {Duarte}, \citenamefont {González-Borrero}, \citenamefont {Lourenço},
  \citenamefont {Mazzaro}, \citenamefont {Marega},\ and\ \citenamefont
  {Salamo}}]{Teodoro2008}%
  \BibitemOpen
  \bibfield  {author} {\bibinfo {author} {\bibfnamefont {M.~D.}\ \bibnamefont
  {Teodoro}}, \bibinfo {author} {\bibfnamefont {I.~F.~L.}\ \bibnamefont
  {Dias}}, \bibinfo {author} {\bibfnamefont {E.}~\bibnamefont {Laureto}},
  \bibinfo {author} {\bibfnamefont {J.~L.}\ \bibnamefont {Duarte}}, \bibinfo
  {author} {\bibfnamefont {P.~P.}\ \bibnamefont {González-Borrero}}, \bibinfo
  {author} {\bibfnamefont {S.~A.}\ \bibnamefont {Lourenço}}, \bibinfo {author}
  {\bibfnamefont {I.}~\bibnamefont {Mazzaro}}, \bibinfo {author} {\bibfnamefont
  {E.}~\bibnamefont {Marega}},\ and\ \bibinfo {author} {\bibfnamefont {G.~J.}\
  \bibnamefont {Salamo}},\ }\bibfield  {title} {\bibinfo {title} {{Substrate
  orientation effect on potential fluctuations in multiquantum wells of
  GaAs/AlGaAs}},\ }\href {https://doi.org/10.1063/1.2913513} {\bibfield
  {journal} {\bibinfo  {journal} {J. Appl. Phys.}\ }\textbf {\bibinfo {volume}
  {103}},\ \bibinfo {pages} {093508} (\bibinfo {year} {2008})}\BibitemShut
  {NoStop}%
\bibitem [{\citenamefont {Runge}(2003)}]{RUNGE2003149}%
  \BibitemOpen
  \bibfield  {author} {\bibinfo {author} {\bibfnamefont {E.}~\bibnamefont
  {Runge}},\ }\bibfield  {title} {\bibinfo {title} {Excitons in semiconductor
  nanostructures}\ }(\bibinfo  {publisher} {Academic Press},\ \bibinfo {year}
  {2003})\ pp.\ \bibinfo {pages} {149--305}\BibitemShut {NoStop}%
\bibitem [{\citenamefont {Mattheis}\ \emph {et~al.}(2007)\citenamefont
  {Mattheis}, \citenamefont {Rau},\ and\ \citenamefont
  {Werner}}]{Mattheis2007}%
  \BibitemOpen
  \bibfield  {author} {\bibinfo {author} {\bibfnamefont {J.}~\bibnamefont
  {Mattheis}}, \bibinfo {author} {\bibfnamefont {U.}~\bibnamefont {Rau}},\ and\
  \bibinfo {author} {\bibfnamefont {J.~H.}\ \bibnamefont {Werner}},\ }\bibfield
   {title} {\bibinfo {title} {{Light absorption and emission in semiconductors
  with band gap fluctuations—A study on
  ${\mathrm{Cu}}({\mathrm{In,Ga}}){\mathrm{Se}}_{2}$ thin films}},\ }\href
  {https://doi.org/10.1063/1.2721768} {\bibfield  {journal} {\bibinfo
  {journal} {J. Appl. Phys.}\ }\textbf {\bibinfo {volume} {101}},\ \bibinfo
  {pages} {113519} (\bibinfo {year} {2007})}\BibitemShut {NoStop}%
\bibitem [{\citenamefont {Christen}\ and\ \citenamefont
  {Bimberg}(1990)}]{Christen1990}%
  \BibitemOpen
  \bibfield  {author} {\bibinfo {author} {\bibfnamefont {J.}~\bibnamefont
  {Christen}}\ and\ \bibinfo {author} {\bibfnamefont {D.}~\bibnamefont
  {Bimberg}},\ }\bibfield  {title} {\bibinfo {title} {{Line shapes of
  intersubband and excitonic recombination in quantum wells: Influence of
  final-state interaction, statistical broadening, and momentum
  conservation}},\ }\href {https://doi.org/10.1103/PhysRevB.42.7213} {\bibfield
   {journal} {\bibinfo  {journal} {Phys. Rev. B}\ }\textbf {\bibinfo {volume}
  {42}},\ \bibinfo {pages} {7213} (\bibinfo {year} {1990})}\BibitemShut
  {NoStop}%
\bibitem [{\citenamefont {Andrews}\ and\ \citenamefont {of~Photo-optical
  Instrumentation~Engineers}(1998)}]{andrews_1997}%
  \BibitemOpen
  \bibfield  {author} {\bibinfo {author} {\bibfnamefont {L.}~\bibnamefont
  {Andrews}}\ and\ \bibinfo {author} {\bibfnamefont {S.}~\bibnamefont
  {of~Photo-optical Instrumentation~Engineers}},\ }\href@noop {} {\emph
  {\bibinfo {title} {Special Functions of Mathematics for Engineers}}},\
  \bibinfo {edition} {2nd}\ ed.,\ Online access with subscription: SPIE Digital
  Library\ (\bibinfo  {publisher} {SPIE Optical Engineering Press},\ \bibinfo
  {address} {Washington},\ \bibinfo {year} {1998})\BibitemShut {NoStop}%
\bibitem [{\citenamefont {Whiting}(1968)}]{Whiting1968}%
  \BibitemOpen
  \bibfield  {author} {\bibinfo {author} {\bibfnamefont {E.}~\bibnamefont
  {Whiting}},\ }\bibfield  {title} {\bibinfo {title} {{An empirical
  approximation to the Voigt profile}},\ }\href
  {https://doi.org/https://doi.org/10.1016/0022-4073(68)90081-2} {\bibfield
  {journal} {\bibinfo  {journal} {J. Quant. Spectrosc. Radiat. Transfer}\
  }\textbf {\bibinfo {volume} {8}},\ \bibinfo {pages} {1379} (\bibinfo {year}
  {1968})}\BibitemShut {NoStop}%
\bibitem [{\citenamefont {Olivero}\ and\ \citenamefont
  {Longbothum}(1977)}]{Olivero1977}%
  \BibitemOpen
  \bibfield  {author} {\bibinfo {author} {\bibfnamefont {J.}~\bibnamefont
  {Olivero}}\ and\ \bibinfo {author} {\bibfnamefont {R.}~\bibnamefont
  {Longbothum}},\ }\bibfield  {title} {\bibinfo {title} {{Empirical fits to the
  Voigt line width: A brief review}},\ }\href
  {https://doi.org/https://doi.org/10.1016/0022-4073(77)90161-3} {\bibfield
  {journal} {\bibinfo  {journal} {J. Quant. Spectrosc. Radiat. Transfer}\
  }\textbf {\bibinfo {volume} {17}},\ \bibinfo {pages} {233} (\bibinfo {year}
  {1977})}\BibitemShut {NoStop}%
\bibitem [{\citenamefont {Lee}\ \emph {et~al.}(1986)\citenamefont {Lee},
  \citenamefont {Koteles},\ and\ \citenamefont {Vassell}}]{lee1986FWHM}%
  \BibitemOpen
  \bibfield  {author} {\bibinfo {author} {\bibfnamefont {J.}~\bibnamefont
  {Lee}}, \bibinfo {author} {\bibfnamefont {E.~S.}\ \bibnamefont {Koteles}},\
  and\ \bibinfo {author} {\bibfnamefont {M.~O.}\ \bibnamefont {Vassell}},\
  }\bibfield  {title} {\bibinfo {title} {{Luminescence linewidths of excitons
  in GaAs quantum wells below 150 K}},\ }\href
  {https://doi.org/10.1103/PhysRevB.33.5512} {\bibfield  {journal} {\bibinfo
  {journal} {Phys. Rev. B}\ }\textbf {\bibinfo {volume} {33}},\ \bibinfo
  {pages} {5512} (\bibinfo {year} {1986})}\BibitemShut {NoStop}%
\bibitem [{\citenamefont {Aksenov}\ \emph {et~al.}(1995)\citenamefont
  {Aksenov}, \citenamefont {Kusano}, \citenamefont {Aoyagi}, \citenamefont
  {Sugano}, \citenamefont {Yasuda},\ and\ \citenamefont
  {Segawa}}]{Aksenov1995}%
  \BibitemOpen
  \bibfield  {author} {\bibinfo {author} {\bibfnamefont {I.}~\bibnamefont
  {Aksenov}}, \bibinfo {author} {\bibfnamefont {J.}~\bibnamefont {Kusano}},
  \bibinfo {author} {\bibfnamefont {Y.}~\bibnamefont {Aoyagi}}, \bibinfo
  {author} {\bibfnamefont {T.}~\bibnamefont {Sugano}}, \bibinfo {author}
  {\bibfnamefont {T.}~\bibnamefont {Yasuda}},\ and\ \bibinfo {author}
  {\bibfnamefont {Y.}~\bibnamefont {Segawa}},\ }\bibfield  {title} {\bibinfo
  {title} {Effect of a magnetic field on the excitonic luminescence line shape
  in a quantum well},\ }\href {https://doi.org/10.1103/PhysRevB.51.4278}
  {\bibfield  {journal} {\bibinfo  {journal} {Phys. Rev. B}\ }\textbf {\bibinfo
  {volume} {51}},\ \bibinfo {pages} {4278} (\bibinfo {year}
  {1995})}\BibitemShut {NoStop}%
\bibitem [{\citenamefont {Bansal}\ \emph {et~al.}(2007)\citenamefont {Bansal},
  \citenamefont {Hayne}, \citenamefont {Arora},\ and\ \citenamefont
  {Moshchalkov}}]{Bansal2007}%
  \BibitemOpen
  \bibfield  {author} {\bibinfo {author} {\bibfnamefont {B.}~\bibnamefont
  {Bansal}}, \bibinfo {author} {\bibfnamefont {M.}~\bibnamefont {Hayne}},
  \bibinfo {author} {\bibfnamefont {B.~M.}\ \bibnamefont {Arora}},\ and\
  \bibinfo {author} {\bibfnamefont {V.~V.}\ \bibnamefont {Moshchalkov}},\
  }\bibfield  {title} {\bibinfo {title} {{Magnetic field-dependent
  photoluminescence linewidths as a probe of disorder length scales in quantum
  wells}},\ }\href {https://doi.org/10.1063/1.2825417} {\bibfield  {journal}
  {\bibinfo  {journal} {Appl. Phys. Lett.}\ }\textbf {\bibinfo {volume} {91}},\
  \bibinfo {pages} {251108} (\bibinfo {year} {2007})}\BibitemShut {NoStop}%
\bibitem [{\citenamefont {Chen}\ \emph {et~al.}(2020)\citenamefont {Chen},
  \citenamefont {Xu}, \citenamefont {Zhou}, \citenamefont {Zhu}, \citenamefont
  {Chen},\ and\ \citenamefont {Shao}}]{Chen2020}%
  \BibitemOpen
  \bibfield  {author} {\bibinfo {author} {\bibfnamefont {X.}~\bibnamefont
  {Chen}}, \bibinfo {author} {\bibfnamefont {Z.}~\bibnamefont {Xu}}, \bibinfo
  {author} {\bibfnamefont {Y.}~\bibnamefont {Zhou}}, \bibinfo {author}
  {\bibfnamefont {L.}~\bibnamefont {Zhu}}, \bibinfo {author} {\bibfnamefont
  {J.}~\bibnamefont {Chen}},\ and\ \bibinfo {author} {\bibfnamefont
  {J.}~\bibnamefont {Shao}},\ }\bibfield  {title} {\bibinfo {title}
  {{Evaluating interface roughness and micro-fluctuation potential of InAs/GaSb
  superlattices by mid-infrared magnetophotoluminescence}},\ }\href
  {https://doi.org/10.1063/5.0015540} {\bibfield  {journal} {\bibinfo
  {journal} {Appl. Phys. Lett.}\ }\textbf {\bibinfo {volume} {117}},\ \bibinfo
  {pages} {081104} (\bibinfo {year} {2020})}\BibitemShut {NoStop}%
\bibitem [{\citenamefont {Glazov}(2004)}]{Glazov}%
  \BibitemOpen
  \bibfield  {author} {\bibinfo {author} {\bibfnamefont {M.~M.}\ \bibnamefont
  {Glazov}},\ }\bibfield  {title} {\bibinfo {title} {{Magnetic field effects on
  spin relaxation in heterostructures}},\ }\href
  {https://doi.org/10.1103/PhysRevB.70.195314} {\bibfield  {journal} {\bibinfo
  {journal} {Phys. Rev. B}\ }\textbf {\bibinfo {volume} {70}},\ \bibinfo
  {pages} {195314} (\bibinfo {year} {2004})}\BibitemShut {NoStop}%
\bibitem [{\citenamefont {Ando}\ and\ \citenamefont {Uemura}(1974)}]{Ando1974}%
  \BibitemOpen
  \bibfield  {author} {\bibinfo {author} {\bibfnamefont {T.}~\bibnamefont
  {Ando}}\ and\ \bibinfo {author} {\bibfnamefont {Y.}~\bibnamefont {Uemura}},\
  }\bibfield  {title} {\bibinfo {title} {{Theory of Quantum Transport in a
  Two-Dimensional Electron System under Magnetic Fields. I. Characteristics of
  Level Broadening and Transport under Strong Fields}},\ }\href
  {https://doi.org/10.1143/JPSJ.36.959} {\bibfield  {journal} {\bibinfo
  {journal} {J. Phys. Soc. Jpn.}\ }\textbf {\bibinfo {volume} {36}},\ \bibinfo
  {pages} {959} (\bibinfo {year} {1974})}\BibitemShut {NoStop}%
\bibitem [{\citenamefont {Butov}\ \emph {et~al.}(1994)\citenamefont {Butov},
  \citenamefont {Zrenner}, \citenamefont {Abstreiter}, \citenamefont {B\"ohm},\
  and\ \citenamefont {Weimann}}]{Butov1994}%
  \BibitemOpen
  \bibfield  {author} {\bibinfo {author} {\bibfnamefont {L.~V.}\ \bibnamefont
  {Butov}}, \bibinfo {author} {\bibfnamefont {A.}~\bibnamefont {Zrenner}},
  \bibinfo {author} {\bibfnamefont {G.}~\bibnamefont {Abstreiter}}, \bibinfo
  {author} {\bibfnamefont {G.}~\bibnamefont {B\"ohm}},\ and\ \bibinfo {author}
  {\bibfnamefont {G.}~\bibnamefont {Weimann}},\ }\bibfield  {title} {\bibinfo
  {title} {{Condensation of Indirect Excitons in Coupled AlAs/GaAs Quantum
  Wells}},\ }\href {https://doi.org/10.1103/PhysRevLett.73.304} {\bibfield
  {journal} {\bibinfo  {journal} {Phys. Rev. Lett.}\ }\textbf {\bibinfo
  {volume} {73}},\ \bibinfo {pages} {304} (\bibinfo {year} {1994})}\BibitemShut
  {NoStop}%
\bibitem [{\citenamefont {Harrison}\ \emph {et~al.}(2001)\citenamefont
  {Harrison} \emph {et~al.}}]{harrison2001}%
  \BibitemOpen
  \bibfield  {author} {\bibinfo {author} {\bibfnamefont {P.}~\bibnamefont
  {Harrison}} \emph {et~al.},\ }\href@noop {} {\emph {\bibinfo {title}
  {{Quantum wells, wires and dots}}}},\ \bibinfo {edition} {4th}\ ed.\
  (\bibinfo  {publisher} {Wiley Online Library},\ \bibinfo {year}
  {2001})\BibitemShut {NoStop}%
\bibitem [{\citenamefont {St\ifmmode~\mbox{\c{e}}\else \c{e}\fi{}pnicki}\ \emph
  {et~al.}(2015{\natexlab{a}})\citenamefont {St\ifmmode~\mbox{\c{e}}\else
  \c{e}\fi{}pnicki}, \citenamefont {Pi\ifmmode~\mbox{\c{e}}\else
  \c{e}\fi{}tka}, \citenamefont {Morier-Genoud}, \citenamefont {Deveaud},\ and\
  \citenamefont {Matuszewski}}]{Stepnicki2015}%
  \BibitemOpen
  \bibfield  {author} {\bibinfo {author} {\bibfnamefont {P.}~\bibnamefont
  {St\ifmmode~\mbox{\c{e}}\else \c{e}\fi{}pnicki}}, \bibinfo {author}
  {\bibfnamefont {B.}~\bibnamefont {Pi\ifmmode~\mbox{\c{e}}\else
  \c{e}\fi{}tka}}, \bibinfo {author} {\bibfnamefont {F.~m.~c.}\ \bibnamefont
  {Morier-Genoud}}, \bibinfo {author} {\bibfnamefont {B.}~\bibnamefont
  {Deveaud}},\ and\ \bibinfo {author} {\bibfnamefont {M.}~\bibnamefont
  {Matuszewski}},\ }\bibfield  {title} {\bibinfo {title} {{Analytical method
  for determining quantum well exciton properties in a magnetic field}},\
  }\href {https://doi.org/10.1103/PhysRevB.91.195302} {\bibfield  {journal}
  {\bibinfo  {journal} {Phys. Rev. B}\ }\textbf {\bibinfo {volume} {91}},\
  \bibinfo {pages} {195302} (\bibinfo {year} {2015}{\natexlab{a}})}\BibitemShut
  {NoStop}%
\bibitem [{\citenamefont {Harrison}\ \emph {et~al.}(2016)\citenamefont
  {Harrison}, \citenamefont {Young}, \citenamefont {Hodgson}, \citenamefont
  {Young}, \citenamefont {Hayne}, \citenamefont {Danos}, \citenamefont
  {Schliwa}, \citenamefont {Strittmatter}, \citenamefont {Lenz}, \citenamefont
  {Eisele}, \citenamefont {Pohl},\ and\ \citenamefont
  {Bimberg}}]{harrison2016}%
  \BibitemOpen
  \bibfield  {author} {\bibinfo {author} {\bibfnamefont {S.}~\bibnamefont
  {Harrison}}, \bibinfo {author} {\bibfnamefont {M.~P.}\ \bibnamefont {Young}},
  \bibinfo {author} {\bibfnamefont {P.~D.}\ \bibnamefont {Hodgson}}, \bibinfo
  {author} {\bibfnamefont {R.~J.}\ \bibnamefont {Young}}, \bibinfo {author}
  {\bibfnamefont {M.}~\bibnamefont {Hayne}}, \bibinfo {author} {\bibfnamefont
  {L.}~\bibnamefont {Danos}}, \bibinfo {author} {\bibfnamefont
  {A.}~\bibnamefont {Schliwa}}, \bibinfo {author} {\bibfnamefont
  {A.}~\bibnamefont {Strittmatter}}, \bibinfo {author} {\bibfnamefont
  {A.}~\bibnamefont {Lenz}}, \bibinfo {author} {\bibfnamefont {H.}~\bibnamefont
  {Eisele}}, \bibinfo {author} {\bibfnamefont {U.~W.}\ \bibnamefont {Pohl}},\
  and\ \bibinfo {author} {\bibfnamefont {D.}~\bibnamefont {Bimberg}},\
  }\bibfield  {title} {\bibinfo {title} {Heterodimensional charge-carrier
  confinement in stacked submonolayer inas in gaas},\ }\href
  {https://doi.org/10.1103/PhysRevB.93.085302} {\bibfield  {journal} {\bibinfo
  {journal} {Phys. Rev. B}\ }\textbf {\bibinfo {volume} {93}},\ \bibinfo
  {pages} {085302} (\bibinfo {year} {2016})}\BibitemShut {NoStop}%
\bibitem [{\citenamefont {Mena}\ \emph {et~al.}(1991)\citenamefont {Mena},
  \citenamefont {Sanders}, \citenamefont {Bajaj},\ and\ \citenamefont
  {Dudley}}]{mena1991}%
  \BibitemOpen
  \bibfield  {author} {\bibinfo {author} {\bibfnamefont {R.~A.}\ \bibnamefont
  {Mena}}, \bibinfo {author} {\bibfnamefont {G.~D.}\ \bibnamefont {Sanders}},
  \bibinfo {author} {\bibfnamefont {K.~K.}\ \bibnamefont {Bajaj}},\ and\
  \bibinfo {author} {\bibfnamefont {S.~C.}\ \bibnamefont {Dudley}},\ }\bibfield
   {title} {\bibinfo {title} {Theory of the effect of magnetic field on the
  excitonic photoluminescence linewidth in semiconductor alloys},\ }\href
  {https://doi.org/10.1063/1.349509} {\bibfield  {journal} {\bibinfo  {journal}
  {Journal of Applied Physics}\ }\textbf {\bibinfo {volume} {70}},\ \bibinfo
  {pages} {1866} (\bibinfo {year} {1991})},\ \Eprint
  {https://arxiv.org/abs/https://doi.org/10.1063/1.349509}
  {https://doi.org/10.1063/1.349509} \BibitemShut {NoStop}%
\bibitem [{\citenamefont {Goede}\ \emph {et~al.}(1978)\citenamefont {Goede},
  \citenamefont {John},\ and\ \citenamefont {Hennig}}]{goede1978}%
  \BibitemOpen
  \bibfield  {author} {\bibinfo {author} {\bibfnamefont {O.}~\bibnamefont
  {Goede}}, \bibinfo {author} {\bibfnamefont {L.}~\bibnamefont {John}},\ and\
  \bibinfo {author} {\bibfnamefont {D.}~\bibnamefont {Hennig}},\ }\bibfield
  {title} {\bibinfo {title} {Compositional disorder-induced broadening for free
  excitons in ii-vi semiconducting mixed crystals},\ }\href
  {https://doi.org/https://doi.org/10.1002/pssb.2220890262} {\bibfield
  {journal} {\bibinfo  {journal} {physica status solidi (b)}\ }\textbf
  {\bibinfo {volume} {89}},\ \bibinfo {pages} {K183} (\bibinfo {year}
  {1978})}\BibitemShut {NoStop}%
\bibitem [{\citenamefont {Singh}\ and\ \citenamefont
  {Bajaj}(1984)}]{singh1984}%
  \BibitemOpen
  \bibfield  {author} {\bibinfo {author} {\bibfnamefont {J.}~\bibnamefont
  {Singh}}\ and\ \bibinfo {author} {\bibfnamefont {K.~K.}\ \bibnamefont
  {Bajaj}},\ }\bibfield  {title} {\bibinfo {title} {Theory of excitonic
  photoluminescence linewidth in semiconductor alloys},\ }\href
  {https://doi.org/10.1063/1.94649} {\bibfield  {journal} {\bibinfo  {journal}
  {Applied Physics Letters}\ }\textbf {\bibinfo {volume} {44}},\ \bibinfo
  {pages} {1075} (\bibinfo {year} {1984})},\ \Eprint
  {https://arxiv.org/abs/https://doi.org/10.1063/1.94649}
  {https://doi.org/10.1063/1.94649} \BibitemShut {NoStop}%
\bibitem [{\citenamefont {Schubert}\ \emph {et~al.}(1984)\citenamefont
  {Schubert}, \citenamefont {G\"obel}, \citenamefont {Horikoshi}, \citenamefont
  {Ploog},\ and\ \citenamefont {Queisser}}]{schubert1984}%
  \BibitemOpen
  \bibfield  {author} {\bibinfo {author} {\bibfnamefont {E.~F.}\ \bibnamefont
  {Schubert}}, \bibinfo {author} {\bibfnamefont {E.~O.}\ \bibnamefont
  {G\"obel}}, \bibinfo {author} {\bibfnamefont {Y.}~\bibnamefont {Horikoshi}},
  \bibinfo {author} {\bibfnamefont {K.}~\bibnamefont {Ploog}},\ and\ \bibinfo
  {author} {\bibfnamefont {H.~J.}\ \bibnamefont {Queisser}},\ }\bibfield
  {title} {\bibinfo {title} {Alloy broadening in photoluminescence spectra of
  ${\mathrm{al}}_{x}{\mathrm{ga}}_{1\ensuremath{-}x}\mathrm{As}$},\ }\href
  {https://doi.org/10.1103/PhysRevB.30.813} {\bibfield  {journal} {\bibinfo
  {journal} {Phys. Rev. B}\ }\textbf {\bibinfo {volume} {30}},\ \bibinfo
  {pages} {813} (\bibinfo {year} {1984})}\BibitemShut {NoStop}%
\bibitem [{\citenamefont {Singh}\ and\ \citenamefont
  {Bajaj}(1986)}]{singh1986}%
  \BibitemOpen
  \bibfield  {author} {\bibinfo {author} {\bibfnamefont {J.}~\bibnamefont
  {Singh}}\ and\ \bibinfo {author} {\bibfnamefont {K.~K.}\ \bibnamefont
  {Bajaj}},\ }\bibfield  {title} {\bibinfo {title} {Quantum mechanical theory
  of linewidths of localized radiative transitions in semiconductor alloys},\
  }\href {https://doi.org/10.1063/1.96602} {\bibfield  {journal} {\bibinfo
  {journal} {Applied Physics Letters}\ }\textbf {\bibinfo {volume} {48}},\
  \bibinfo {pages} {1077} (\bibinfo {year} {1986})},\ \Eprint
  {https://arxiv.org/abs/https://doi.org/10.1063/1.96602}
  {https://doi.org/10.1063/1.96602} \BibitemShut {NoStop}%
\bibitem [{\citenamefont {Sakaki}\ \emph {et~al.}(1985)\citenamefont {Sakaki},
  \citenamefont {Arakawa}, \citenamefont {Nishioka}, \citenamefont {Yoshino},
  \citenamefont {Okamoto},\ and\ \citenamefont {Miura}}]{sakaki1985}%
  \BibitemOpen
  \bibfield  {author} {\bibinfo {author} {\bibfnamefont {H.}~\bibnamefont
  {Sakaki}}, \bibinfo {author} {\bibfnamefont {Y.}~\bibnamefont {Arakawa}},
  \bibinfo {author} {\bibfnamefont {M.}~\bibnamefont {Nishioka}}, \bibinfo
  {author} {\bibfnamefont {J.}~\bibnamefont {Yoshino}}, \bibinfo {author}
  {\bibfnamefont {H.}~\bibnamefont {Okamoto}},\ and\ \bibinfo {author}
  {\bibfnamefont {N.}~\bibnamefont {Miura}},\ }\bibfield  {title} {\bibinfo
  {title} {Light emission from zero‐dimensional excitons—photoluminescence
  from quantum wells in strong magnetic fields},\ }\href
  {https://doi.org/10.1063/1.95806} {\bibfield  {journal} {\bibinfo  {journal}
  {Applied Physics Letters}\ }\textbf {\bibinfo {volume} {46}},\ \bibinfo
  {pages} {83} (\bibinfo {year} {1985})},\ \Eprint
  {https://arxiv.org/abs/https://doi.org/10.1063/1.95806}
  {https://doi.org/10.1063/1.95806} \BibitemShut {NoStop}%
\bibitem [{\citenamefont {Vahala}\ \emph {et~al.}(1987)\citenamefont {Vahala},
  \citenamefont {Arakawa},\ and\ \citenamefont {Yariv}}]{vahala1987}%
  \BibitemOpen
  \bibfield  {author} {\bibinfo {author} {\bibfnamefont {K.}~\bibnamefont
  {Vahala}}, \bibinfo {author} {\bibfnamefont {Y.}~\bibnamefont {Arakawa}},\
  and\ \bibinfo {author} {\bibfnamefont {A.}~\bibnamefont {Yariv}},\ }\bibfield
   {title} {\bibinfo {title} {Reduction of the field spectrum linewidth of a
  multiple quantum well laser in a high magnetic field—spectral properties of
  quantum dot lasers},\ }\href {https://doi.org/10.1063/1.98200} {\bibfield
  {journal} {\bibinfo  {journal} {Applied Physics Letters}\ }\textbf {\bibinfo
  {volume} {50}},\ \bibinfo {pages} {365} (\bibinfo {year} {1987})},\ \Eprint
  {https://arxiv.org/abs/https://doi.org/10.1063/1.98200}
  {https://doi.org/10.1063/1.98200} \BibitemShut {NoStop}%
\bibitem [{\citenamefont {Oliveira}\ \emph {et~al.}(1999)\citenamefont
  {Oliveira}, \citenamefont {Meneses},\ and\ \citenamefont
  {Silva}}]{oliveira1999}%
  \BibitemOpen
  \bibfield  {author} {\bibinfo {author} {\bibfnamefont {J.~B. B.~d.}\
  \bibnamefont {Oliveira}}, \bibinfo {author} {\bibfnamefont {E.~A.}\
  \bibnamefont {Meneses}},\ and\ \bibinfo {author} {\bibfnamefont {E.~C.
  F.~d.}\ \bibnamefont {Silva}},\ }\bibfield  {title} {\bibinfo {title}
  {Magneto-optical studies of the correlation between interface microroughness
  parameters and the photoluminescence line shape in
  $\mathrm{GaAs}/{\mathrm{ga}}_{0.7}{\mathrm{al}}_{0.3}\mathrm{As}$ quantum
  wells},\ }\href {https://doi.org/10.1103/PhysRevB.60.1519} {\bibfield
  {journal} {\bibinfo  {journal} {Phys. Rev. B}\ }\textbf {\bibinfo {volume}
  {60}},\ \bibinfo {pages} {1519} (\bibinfo {year} {1999})}\BibitemShut
  {NoStop}%
\bibitem [{\citenamefont {Polimeni}\ \emph {et~al.}(2002)\citenamefont
  {Polimeni}, \citenamefont {Patanè}, \citenamefont {Hayden}, \citenamefont
  {Eaves}, \citenamefont {Henini}, \citenamefont {Main}, \citenamefont
  {Uchida}, \citenamefont {Miura}, \citenamefont {Main},\ and\ \citenamefont
  {Wunner}}]{POLIMENI2002}%
  \BibitemOpen
  \bibfield  {author} {\bibinfo {author} {\bibfnamefont {A.}~\bibnamefont
  {Polimeni}}, \bibinfo {author} {\bibfnamefont {A.}~\bibnamefont {Patanè}},
  \bibinfo {author} {\bibfnamefont {R.}~\bibnamefont {Hayden}}, \bibinfo
  {author} {\bibfnamefont {L.}~\bibnamefont {Eaves}}, \bibinfo {author}
  {\bibfnamefont {M.}~\bibnamefont {Henini}}, \bibinfo {author} {\bibfnamefont
  {P.}~\bibnamefont {Main}}, \bibinfo {author} {\bibfnamefont {K.}~\bibnamefont
  {Uchida}}, \bibinfo {author} {\bibfnamefont {N.}~\bibnamefont {Miura}},
  \bibinfo {author} {\bibfnamefont {J.}~\bibnamefont {Main}},\ and\ \bibinfo
  {author} {\bibfnamefont {G.}~\bibnamefont {Wunner}},\ }\bibfield  {title}
  {\bibinfo {title} {Linewidth broadening of excitonic luminescence from
  quantum wells in pulsed magnetic fields},\ }\href
  {https://doi.org/https://doi.org/10.1016/S1386-9477(01)00555-0} {\bibfield
  {journal} {\bibinfo  {journal} {Physica E: Low-dimensional Systems and
  Nanostructures}\ }\textbf {\bibinfo {volume} {13}},\ \bibinfo {pages} {349}
  (\bibinfo {year} {2002})}\BibitemShut {NoStop}%
\bibitem [{\citenamefont {St\ifmmode~\mbox{\c{e}}\else \c{e}\fi{}pnicki}\ \emph
  {et~al.}(2015{\natexlab{b}})\citenamefont {St\ifmmode~\mbox{\c{e}}\else
  \c{e}\fi{}pnicki}, \citenamefont {Pi\ifmmode~\mbox{\c{e}}\else
  \c{e}\fi{}tka}, \citenamefont {Morier-Genoud}, \citenamefont {Deveaud},\ and\
  \citenamefont {Matuszewski}}]{piotr2015}%
  \BibitemOpen
  \bibfield  {author} {\bibinfo {author} {\bibfnamefont {P.}~\bibnamefont
  {St\ifmmode~\mbox{\c{e}}\else \c{e}\fi{}pnicki}}, \bibinfo {author}
  {\bibfnamefont {B.}~\bibnamefont {Pi\ifmmode~\mbox{\c{e}}\else
  \c{e}\fi{}tka}}, \bibinfo {author} {\bibfnamefont {F.~m.~c.}\ \bibnamefont
  {Morier-Genoud}}, \bibinfo {author} {\bibfnamefont {B.}~\bibnamefont
  {Deveaud}},\ and\ \bibinfo {author} {\bibfnamefont {M.}~\bibnamefont
  {Matuszewski}},\ }\bibfield  {title} {\bibinfo {title} {Analytical method for
  determining quantum well exciton properties in a magnetic field},\ }\href
  {https://doi.org/10.1103/PhysRevB.91.195302} {\bibfield  {journal} {\bibinfo
  {journal} {Phys. Rev. B}\ }\textbf {\bibinfo {volume} {91}},\ \bibinfo
  {pages} {195302} (\bibinfo {year} {2015}{\natexlab{b}})}\BibitemShut
  {NoStop}%
\bibitem [{\citenamefont {Wilamowski}\ and\ \citenamefont
  {Jantsch}(2004)}]{wilam2004}%
  \BibitemOpen
  \bibfield  {author} {\bibinfo {author} {\bibfnamefont {Z.}~\bibnamefont
  {Wilamowski}}\ and\ \bibinfo {author} {\bibfnamefont {W.}~\bibnamefont
  {Jantsch}},\ }\bibfield  {title} {\bibinfo {title} {Suppression of spin
  relaxation of conduction electrons by cyclotron motion},\ }\href
  {https://doi.org/10.1103/PhysRevB.69.035328} {\bibfield  {journal} {\bibinfo
  {journal} {Phys. Rev. B}\ }\textbf {\bibinfo {volume} {69}},\ \bibinfo
  {pages} {035328} (\bibinfo {year} {2004})}\BibitemShut {NoStop}%
\bibitem [{\citenamefont {Llorens}\ \emph {et~al.}(2019)\citenamefont
  {Llorens}, \citenamefont {Lopes-Oliveira}, \citenamefont {L\'opez-Richard},
  \citenamefont {de~Oliveira}, \citenamefont {Wewi\'or}, \citenamefont {Ulloa},
  \citenamefont {Teodoro}, \citenamefont {Marques}, \citenamefont
  {Garc\'{\i}a-Crist\'obal}, \citenamefont {Hai},\ and\ \citenamefont
  {Al\'en}}]{llorens2019}%
  \BibitemOpen
  \bibfield  {author} {\bibinfo {author} {\bibfnamefont {J.}~\bibnamefont
  {Llorens}}, \bibinfo {author} {\bibfnamefont {V.}~\bibnamefont
  {Lopes-Oliveira}}, \bibinfo {author} {\bibfnamefont {V.}~\bibnamefont
  {L\'opez-Richard}}, \bibinfo {author} {\bibfnamefont {E.~C.}\ \bibnamefont
  {de~Oliveira}}, \bibinfo {author} {\bibfnamefont {L.}~\bibnamefont
  {Wewi\'or}}, \bibinfo {author} {\bibfnamefont {J.}~\bibnamefont {Ulloa}},
  \bibinfo {author} {\bibfnamefont {M.}~\bibnamefont {Teodoro}}, \bibinfo
  {author} {\bibfnamefont {G.}~\bibnamefont {Marques}}, \bibinfo {author}
  {\bibfnamefont {A.}~\bibnamefont {Garc\'{\i}a-Crist\'obal}}, \bibinfo
  {author} {\bibfnamefont {G.-Q.}\ \bibnamefont {Hai}},\ and\ \bibinfo {author}
  {\bibfnamefont {B.}~\bibnamefont {Al\'en}},\ }\bibfield  {title} {\bibinfo
  {title} {Topology driven $g$-factor tuning in type-ii quantum dots},\ }\href
  {https://doi.org/10.1103/PhysRevApplied.11.044011} {\bibfield  {journal}
  {\bibinfo  {journal} {Phys. Rev. Applied}\ }\textbf {\bibinfo {volume}
  {11}},\ \bibinfo {pages} {044011} (\bibinfo {year} {2019})}\BibitemShut
  {NoStop}%
\bibitem [{\citenamefont {Huant}\ \emph {et~al.}(1992)\citenamefont {Huant},
  \citenamefont {Mandray},\ and\ \citenamefont {Etienne}}]{huant1992}%
  \BibitemOpen
  \bibfield  {author} {\bibinfo {author} {\bibfnamefont {S.}~\bibnamefont
  {Huant}}, \bibinfo {author} {\bibfnamefont {A.}~\bibnamefont {Mandray}},\
  and\ \bibinfo {author} {\bibfnamefont {B.}~\bibnamefont {Etienne}},\
  }\bibfield  {title} {\bibinfo {title} {Nonparabolicity effects on cyclotron
  mass in gaas quantum wells},\ }\href
  {https://doi.org/10.1103/PhysRevB.46.2613} {\bibfield  {journal} {\bibinfo
  {journal} {Phys. Rev. B}\ }\textbf {\bibinfo {volume} {46}},\ \bibinfo
  {pages} {2613} (\bibinfo {year} {1992})}\BibitemShut {NoStop}%
\bibitem [{\citenamefont {Winkler}\ \emph {et~al.}(2003)\citenamefont
  {Winkler}, \citenamefont {Papadakis}, \citenamefont {De~Poortere},\ and\
  \citenamefont {Shayegan}}]{winkler2003}%
  \BibitemOpen
  \bibfield  {author} {\bibinfo {author} {\bibfnamefont {R.}~\bibnamefont
  {Winkler}}, \bibinfo {author} {\bibfnamefont {S.}~\bibnamefont {Papadakis}},
  \bibinfo {author} {\bibfnamefont {E.}~\bibnamefont {De~Poortere}},\ and\
  \bibinfo {author} {\bibfnamefont {M.}~\bibnamefont {Shayegan}},\ }\href@noop
  {} {\emph {\bibinfo {title} {Spin-Orbit Coupling in Two-Dimensional Electron
  and Hole Systems}}},\ Vol.~\bibinfo {volume} {41}\ (\bibinfo  {publisher}
  {Springer},\ \bibinfo {year} {2003})\BibitemShut {NoStop}%
\bibitem [{\citenamefont {Skolnick}\ \emph {et~al.}(1976)\citenamefont
  {Skolnick}, \citenamefont {Jain}, \citenamefont {Stradling}, \citenamefont
  {Leotin},\ and\ \citenamefont {Ousset}}]{Skolnick1976}%
  \BibitemOpen
  \bibfield  {author} {\bibinfo {author} {\bibfnamefont {M.~S.}\ \bibnamefont
  {Skolnick}}, \bibinfo {author} {\bibfnamefont {A.~K.}\ \bibnamefont {Jain}},
  \bibinfo {author} {\bibfnamefont {R.~A.}\ \bibnamefont {Stradling}}, \bibinfo
  {author} {\bibfnamefont {J.}~\bibnamefont {Leotin}},\ and\ \bibinfo {author}
  {\bibfnamefont {J.~C.}\ \bibnamefont {Ousset}},\ }\bibfield  {title}
  {\bibinfo {title} {An investigation of the anisotropy of the valence band of
  {GaAs} by cyclotron resonance},\ }\href
  {https://doi.org/10.1088/0022-3719/9/14/019} {\bibfield  {journal} {\bibinfo
  {journal} {Journal of Physics C: Solid State Physics}\ }\textbf {\bibinfo
  {volume} {9}},\ \bibinfo {pages} {2809} (\bibinfo {year} {1976})}\BibitemShut
  {NoStop}%
\bibitem [{\citenamefont {Snelling}\ \emph {et~al.}(1991)\citenamefont
  {Snelling}, \citenamefont {Flinn}, \citenamefont {Plaut}, \citenamefont
  {Harley}, \citenamefont {Tropper}, \citenamefont {Eccleston},\ and\
  \citenamefont {Phillips}}]{snelling1991}%
  \BibitemOpen
  \bibfield  {author} {\bibinfo {author} {\bibfnamefont {M.~J.}\ \bibnamefont
  {Snelling}}, \bibinfo {author} {\bibfnamefont {G.~P.}\ \bibnamefont {Flinn}},
  \bibinfo {author} {\bibfnamefont {A.~S.}\ \bibnamefont {Plaut}}, \bibinfo
  {author} {\bibfnamefont {R.~T.}\ \bibnamefont {Harley}}, \bibinfo {author}
  {\bibfnamefont {A.~C.}\ \bibnamefont {Tropper}}, \bibinfo {author}
  {\bibfnamefont {R.}~\bibnamefont {Eccleston}},\ and\ \bibinfo {author}
  {\bibfnamefont {C.~C.}\ \bibnamefont {Phillips}},\ }\bibfield  {title}
  {\bibinfo {title} {Magnetic g factor of electrons in
  gaas/${\mathrm{al}}_{\mathit{x}}$${\mathrm{ga}}_{1\mathrm{\ensuremath{-}}\mathit{x}}$as
  quantum wells},\ }\href {https://doi.org/10.1103/PhysRevB.44.11345}
  {\bibfield  {journal} {\bibinfo  {journal} {Phys. Rev. B}\ }\textbf {\bibinfo
  {volume} {44}},\ \bibinfo {pages} {11345} (\bibinfo {year}
  {1991})}\BibitemShut {NoStop}%
\bibitem [{\citenamefont {Colocci}\ \emph {et~al.}(1990)\citenamefont
  {Colocci}, \citenamefont {Gurioli}, \citenamefont {Vinattieri}, \citenamefont
  {Fermi}, \citenamefont {Deparis}, \citenamefont {Massies},\ and\
  \citenamefont {Neu}}]{Colocci1990}%
  \BibitemOpen
  \bibfield  {author} {\bibinfo {author} {\bibfnamefont {M.}~\bibnamefont
  {Colocci}}, \bibinfo {author} {\bibfnamefont {M.}~\bibnamefont {Gurioli}},
  \bibinfo {author} {\bibfnamefont {A.}~\bibnamefont {Vinattieri}}, \bibinfo
  {author} {\bibfnamefont {F.}~\bibnamefont {Fermi}}, \bibinfo {author}
  {\bibfnamefont {C.}~\bibnamefont {Deparis}}, \bibinfo {author} {\bibfnamefont
  {J.}~\bibnamefont {Massies}},\ and\ \bibinfo {author} {\bibfnamefont
  {G.}~\bibnamefont {Neu}},\ }\bibfield  {title} {\bibinfo {title} {Temperature
  dependence of exciton lifetimes in {GaAs}/{AlGaAs} quantum well structures},\
  }\href {https://doi.org/10.1209/0295-5075/12/5/007} {\bibfield  {journal}
  {\bibinfo  {journal} {Europhysics Letters ({EPL})}\ }\textbf {\bibinfo
  {volume} {12}},\ \bibinfo {pages} {417} (\bibinfo {year} {1990})}\BibitemShut
  {NoStop}%
\end{thebibliography}%
	
\end{document}